\documentclass[proceedings]{JHEP} 
\usepackage{epsfig}

\def\nubar{\overline{\nu}}
\def\lsim{\stackrel{<}{\sim}}
\def\gsim{\stackrel{>}{\sim}}
\def\lra{\leftrightarrow}
\def\CPv{\not\!\!\!{\rm CP}}
\def\etc{{\it etc.}}
\def\eg{{\it e.g.}}
\def\etal{{\it et al.}}
\def\ie{{\it i.e.}}
\def\BR{{\cal B}}
\def\calA{{\cal A}}
\let\q=\quad

\title{Tau Electroweak Couplings}

\author{Alan J.~Weinstein\thanks{For the CLEO Collaboration.
        Work supported by US DOE Grant DE-FG03-92-ER40701.} \\
        California Institute of Technology \\
        E-mail: \email{ajw@caltech.edu}}

\conference{Heavy Flavours 8, Southampton, UK, 1999}

\abstract{We review world-average measurements of the tau lepton
electroweak couplings, in both decay (including Michel parameters)
and in production ($Z^0\to \tau^+\tau^-$ and $W^-\to\tau^-\nubar_\tau$).
We review the searches for anomalous weak and EM dipole couplings.
Finally, we present the status of several other tau lepton studies:
searches for lepton flavor violating decays, 
neutrino oscillations,
and tau neutrino mass limits.
}

\begin{document} 

\section{Introduction}

Most talks at this conference concern the study of heavy quarks,
and many focus on the difficulties associated with the measurement
of their electroweak couplings, due to their their strong interactions.
This contribution instead focuses on the one heavy flavor fermion
whose electroweak couplings can be measured without such difficulties:
the tau lepton. Indeed, the tau's electroweak couplings
have now been measured with rather high precision and generality,
in both production and decay. In all cases, the couplings of the tau
are identical, to high precision, to those of the electron and the muon.
The leptonic couplings thus form
a standard against which the hypothesis of universality of
all fermionic (including quark) couplings can be tested.

Because the tau lepton is so massive, it decays in many
different ways. The daughter decay products can be
used to analyze the spin polarization of the parent tau.
This can then be used to study the spin dependence of the 
tau electroweak couplings.
Further, since the tau is the heaviest known lepton
and a member of the third family of fermions,
it may be expected to be more sensitive to physics
beyond the Standard Model (SM), especially to
mass-dependent (Higgs-like) currents.
This will reveal itself in violations
of universality of the fermionic couplings.

Here we review the status of the measurements of the tau
electroweak couplings in both production and decay.
The following topics will be covered (necessarily, briefly,
with little attention to experimental detail).
We discuss the tau lifetime, the leptonic branching fractions
($\tau\to e\nu\nu$ and $\mu\nu\nu$), and the results for tests
of universality in the charged current decay.
We then turn to measurements of the Michel Parameters,
which probe deviations from the pure $V-A$ structure of 
the charged weak current.
Next we review the charged current couplings in tau production
via $W^-\to \tau^-\nubar_\tau$ decay.
Then we turn to neutral current couplings in tau production
via $Z^0\to\tau^+\tau^-$.
We review searches for anomalous weak dipole moment
couplings in $Z^0 \to \tau^+ \tau^-$,
and anomalous electromagnetic dipole moment
couplings in $Z^0 \to \tau^+ \tau^-\gamma$.
We briefly review the searches for 
flavor changing neutral currents in
lepton flavor violating (neutrinoless) tau decays,
and searches for neutrino 
oscillations involving the tau neutrino $\nu_\tau$.
We present the current limits on the $\nu_\tau$ mass.
Finally, we summarize and review the prospects for 
further progress in $\tau$ physics in the coming years.

Most of these high-precision measurements and sensitive
searches for anomalous (non-SM) couplings
have been performed, and refined, over the last few years.
There have been no dramatic new results since the last 
Heavy Flavors conference, only updated results with higher precision.
There are updated leptonic branching fractions from LEP;
updated Michel parameter measurements from LEP and CLEO;
final results on tau polarization measurements
and measurements of the $Z^0$ couplings from LEP;
limits on neutral weak dipole moments from LEP and SLD;
new measurements of the rate for $W\to \tau\nu$ from LEP II,
results on electromagnetic dipole moments from LEP,
new $m(\nu_\tau)$ limits from CLEO; and
new limits on lepton flavor violating (neutrinoless)
decays from CLEO.
I draw heavily from the presentations at the Fifth Workshop
on Tau Lepton Physics (TAU'98), from September 1998.

Other than neutrino oscillation observations
which may involve $\nu_\mu\to\nu_\tau$ oscillations,
all recent measurements confirm the minimal 
Standard Model predictions to ever increasing precision.
Nevertheless, new physics may be just around the corner,
waiting to be revealed by even higher precision studies.

\section{Leptonic branching fractions, tau lifetime,
universality}
\label{s-univ}

The rate for tau decays to leptons is given by
the universal charged weak current 
decay rate formula for pointlike massive fermions:


\begin{equation}
\Gamma(\tau\to\nu_\tau \mu \nu_\mu) =
   {\tau_\tau  \BR(\tau\to\nu_\tau \mu \nu_\mu)} \nonumber
\label{eqn:Gtau}
\end{equation}
\begin{equation}
\hspace*{1cm} =
{{{G_F^2 g^2_\tau  g^2_\mu}{m_{\tau}^5}}\over{192\pi^3}} f_{\mu\tau} R_{EW}
   h_\eta
\label{eqn:Gtau2}
\end{equation}
Here, the Fermi coupling constant $G_F$ is {\it measured} 
in $\mu$ decay, assuming $g_\mu g_e$ is 1. 
Here, we let them vary, in order to test the assumption of universality.
The phase space correction $f_{\mu\tau} = f\left(m_\mu^2/m_\tau^2\right)$ 
is 0.9726 for $\tau\to\mu\nu\nu$ and $\approx 1$ for $\tau\to e\nu\nu$.
The electroweak correction is
\begin{eqnarray}
R_{EW} &=& \left(1 + \frac{3}{5}\frac{m_\tau^2}{m^2_W}\right)
   \left[ 1+\frac{\alpha}{\pi}\left(\frac{25}{4} - \pi^2\right)\right]
  \nonumber  \\
   && \hspace*{0.6cm} +0.03\% \hspace*{1cm} -0.4\%
\label{eqn:REW}
\end{eqnarray}
and the correction due to possible scalar currents
is, in terms of the Michel parameter $\eta$,
\begin{eqnarray}
h_\eta = 1/(1+4\eta m_\ell/m_\tau) \q\q \hbox{= 1 in SM.}
\end{eqnarray}

To test the hypothesis that all of the charged weak current couplings
are equal ($g_e = g_\mu = g_\tau$), we must measure the muon lifetime,
the branching fraction $\BR(\mu\to e\nu\nubar)$,
the tau lifetime, and the branching fractions $\BR(\tau\to e\nu\nubar)$
and $\BR(\tau\to \mu\nu\nubar)$. The muon properties are well
measured~\cite{ref:PDG98}.

\subsection{Tau Lifetime}

The tau lifetime has been measured by many experiments with many
different methods~\cite{ref:wasser}.
There are recent measurements from L3~\cite{ref:tautauL3}
and DELPHI~\cite{ref:tautauDELPHI}.

An example of a decay length distribution,
from DELPHI~\cite{ref:tautauDELPHI},
is shown in Fig.~\ref{fig:tautauDELPHI}.
A summary of recent measurements is given in
Fig.~\ref{fig:tautausum}.
The world average from 6 experiments, each
with $\lsim 1\%$ precision, is
$\tau_\tau = (290.5\pm 1.0)$ fs.

\FIGURE[!htb]{
\epsfig{figure=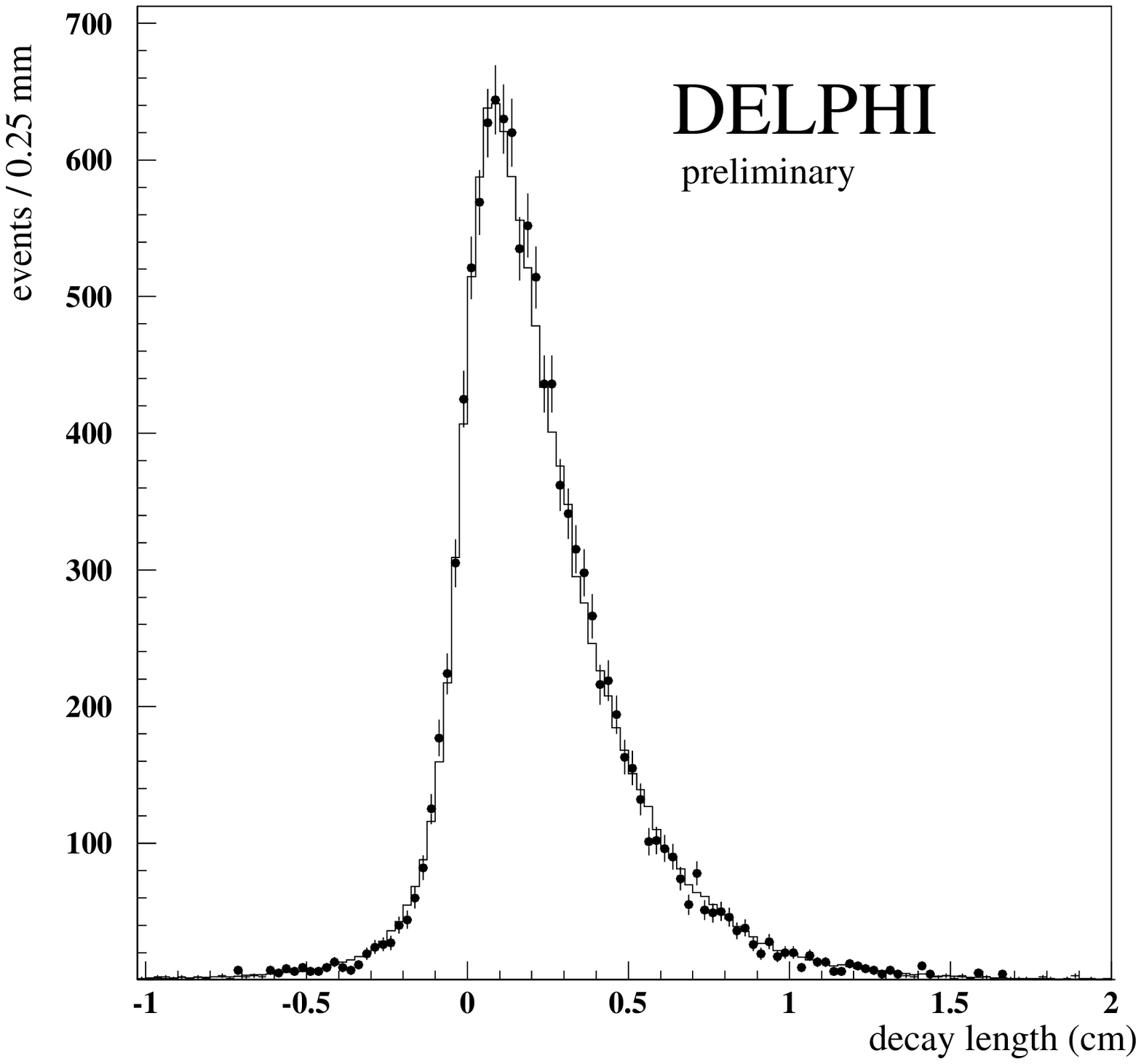,width=4.00cm}
\caption{Tau flight length distribution from DELPHI
{\protect\cite{ref:tautauDELPHI}}.}
\label{fig:tautauDELPHI}
}

\FIGURE[!htb]{
\epsfig{figure=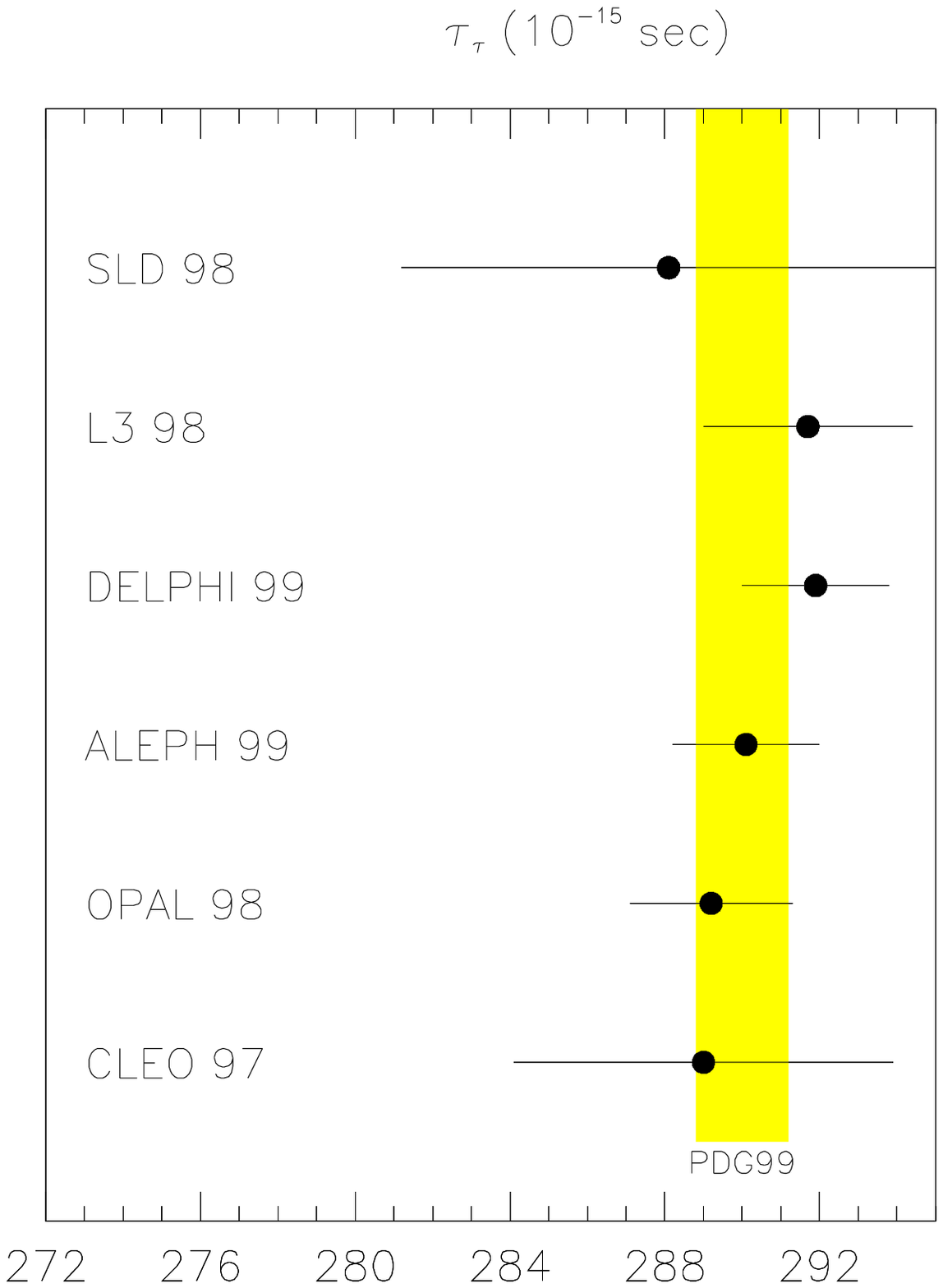,width=5.00cm}
\caption{Summary of recent tau lifetime measurements.}
\label{fig:tautausum}
}

\subsection{Leptonic Branching Fractions}

There are recent results on the tau 
leptonic branching fractions from 
ALEPH~\cite{ref:tauBRALEPH} 
DELPHI~\cite{ref:tauBRDELPHI},
and OPAL~\cite{ref:tauBROPAL}.
Measurements from 5 experiments~\cite{ref:tauBRsum}
are shown in
Fig.~\ref{fig:tauBRsum},
leading to the world average values:
\begin{eqnarray}
\BR(\tau^-\to e^-\nubar_e\nu_\tau)
          &=& (17.81\pm0.07)\% \\
\BR(\tau^-\to \mu^-\nubar_\mu\nu_\tau)
          &=& (17.37\pm0.09)\% 
\end{eqnarray}

\FIGURE[!htb]{
\epsfig{figure=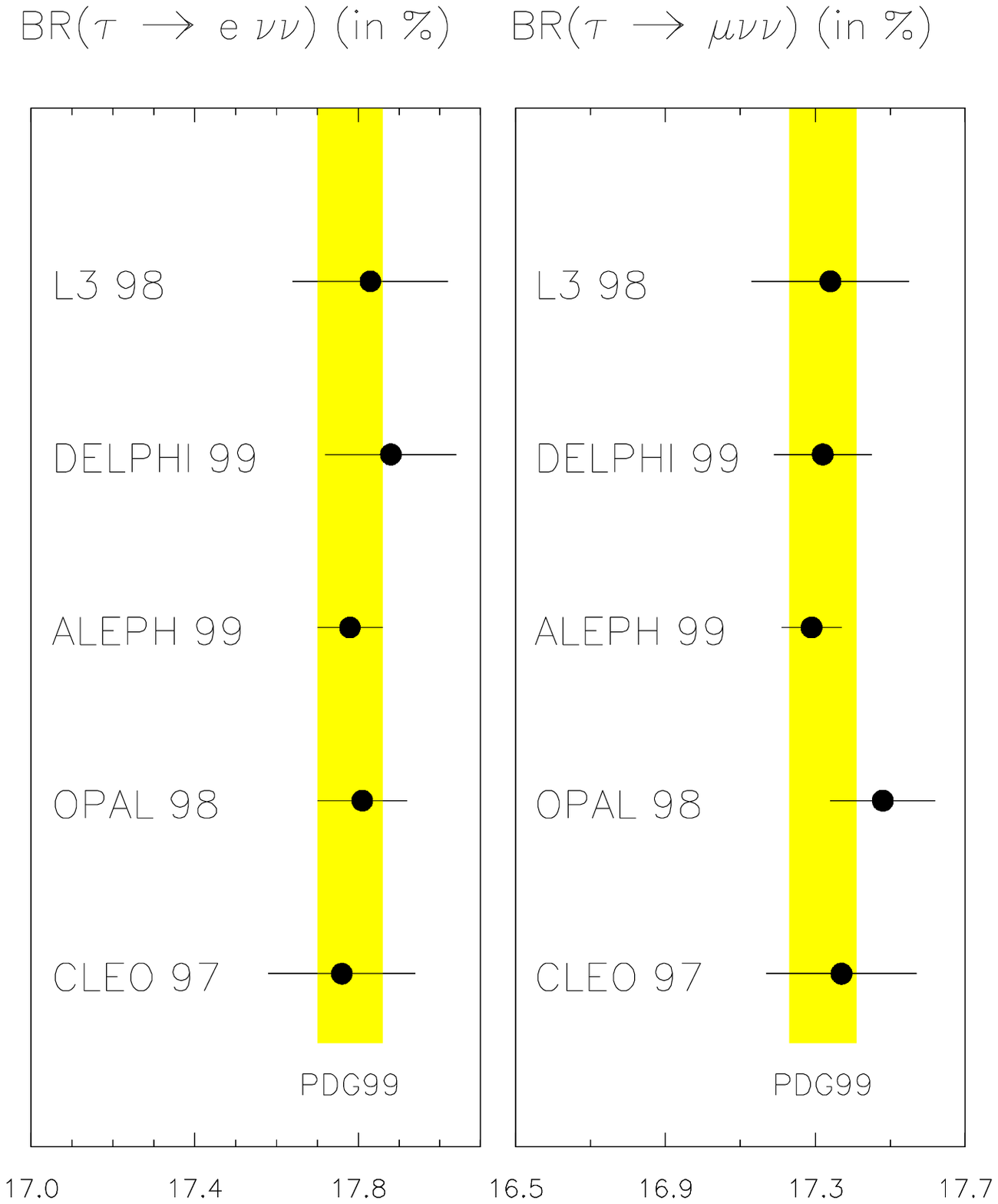,width=6.0cm}
\caption{Summary of recent 
tau leptonic branching fraction measurements.}
\label{fig:tauBRsum}
}

The branching fractions $\BR_e$ and $\BR_\mu$ 
are now measured to 0.4\%\ accuracy,
\ie, at the level of the radiative corrections
(Eqn.~\ref{eqn:REW}).

These precise results already provide
limits on simple extensions to the Standard Model,
using Eqn.~\ref{eqn:Gtau}.
The $\eta$ parameter (to be discussed in section~\ref{s:michel} below)
is inferred to be
\begin{equation}
\eta = 0.013\pm 0.022 \q \hbox{($\eta = 0$ in SM).}
\end{equation}
In addition,
The $\nu_\tau$ mass must be less than 38 MeV~\cite{ref:SwainTaylor}.
One can also put limits on mixing with a $4^{th}$ generation, 
anomalous electromagnetic complings, 
and compositeness~\cite{ref:SwainTaylor}.

\subsection{$\BR(\tau\to\ell\nu\nu\gamma)$ from CLEO 99}

CLEO has made precision measurements of tau leptonic decays
in the presence of a radiative (decay) photon~\cite{ref:tauradCLEO},
finding branching fractions in good agreement with
Standard Model predictions:
\begin{eqnarray}
\BR(e^-\nubar_e\nu_\tau\gamma)
          &=&  (1.75\pm0.06\pm0.17)\% \nonumber \\
     SM   &=& (1.86\pm0.01)\% \\
\BR(\mu^-\nubar_\mu\nu_\tau\gamma)
          &=&  (0.361\pm0.016\pm0.035)\% \nonumber \\
     SM   &=& (0.368\pm0.002)\%
\end{eqnarray}
for $E_\gamma^* > 10$ MeV in the $\tau$ center of mass.

\subsection{Lepton Universality}

From the measurements of the tau lifetime
and leptonic branching fractions,
we can extract ratios which test
the universality hypothesis
$g_e = g_\mu = g_\tau$:

\begin{eqnarray}
\left(\frac{g_\tau}{g_\mu}\right)^2 &\equiv&   \BR_e 
   \left(\frac{\tau_\mu}{\tau_\tau}\right) 
   \left(\frac{m_\mu}{m_\tau}\right)^5 \nonumber \\
   &=&  (1.000\pm0.003)^2 \\
\left(\frac{g_\tau}{g_e}\right)^2 &\equiv&   \frac{\BR_\mu}{f_{\mu\tau}}
   \left(\frac{\tau_\mu}{\tau_\tau}\right)
   \left(\frac{m_\mu}{m_\tau}\right)^5 \nonumber \\
    &=&  (1.000\pm0.003)^2 \\
\left(\frac{g_\mu}{g_e}\right)^2 &\equiv&
   \frac{\BR_\mu}{f_{\mu\tau}\BR_e} \nonumber \\
  &=&  (1.000\pm0.003)^2 
\end{eqnarray}

We see that the {\it strength} of the charged current couplings
(irrespective of their Lorentz structure)
are equal to 0.25\%.


\subsection{What Could Cause Lepton Universality Violation?}

Many extensions to the Standard Model predict violation
of lepton universality.
In fact, lepton universality is put in to the SM by hand,
so that non-universal $W\to\ell\nu$ couplings
can naturally appear if the model is not so constrained.

In the Minimal Supersymmetric SM (MSSM),
decays via a charged Higgs (which couples more strongly
to the heavy tau than to the lighter leptons)
can interfere con- or destructively with the $W$ graph,
enhancing or suppressing the decay rate~\cite{ref:Stahl}.

If a fourth-generation massive $\nu_4$ or sterile $\nu_s$
exists, and mixes with $\nu_\tau$,
it will suppress {\it all} the decay rates and thus
the total tau lifetime.


The current accuracy of the measurements 
do not yield significant limits on any of these 
models, illustrating the need to further improve
their precision.

\section{Michel Parameters}
\label{s:michel}

Although the strength of the charged weak interaction
in decays of the muon and tau are well measured,
the Lorentz structure in tau decays is not as well
established as it is for the purely $V-A$ structure
seen in muon decays.
In general, the couplings can have scalar (S), pseudoscalar (P),
and tensor (T) terms as well as the 
vector (V) and axialvector (A) contributions built into the SM.
Using a general ansatz for the couplings,
including all the lowest-order S, P, V, A, T terms,
Michel~\cite{ref:Michel} derived a form for the 
differential decay rate of the muon (and the tau),
integrating over the unobserved
$\nu$ momenta and daughter $\ell^{\pm}$ spin.
In terms of scaled energy $x=E_{\ell}/E_{max}$,
with $E_{max} = (m_{\ell}^{2}+m_{\tau}^{2})/2m_{\tau}$,
one has:
{\footnotesize
\begin{eqnarray}
\frac{1}{\Gamma} \frac{d\Gamma}{dxd\cos\theta} 
     & = & \frac{x^2}{2} \left[  
                 \left( 12(1-x) + \frac{4{ \rho}}{3}(8x-6) 
                 \right.\right.\nonumber \\
     & &  \left.
           + 24{ \eta}{ \frac{m_\ell}{m_\tau}}
                \frac{(1-x)}{x}\right) \nonumber \\
     &   & \left. - { \xi} { \cos\theta} \left( 4(1-x)+ 
        \frac{4}{3}{ \delta}(8x-6) \right)\right] \\
    & \propto & x^2\left[ I(x\vert {\rho , \eta} ) \pm A( x,{ \theta} \vert
            {\xi ,\delta}) \right]
\end{eqnarray}
}

The spectral shape Michel parameters,
and their SM ($V-A$) values, are:
\begin{eqnarray}
  { \rho}  &\simeq& \frac{3}{4}\left( \frac{ {|g^{V}_{LL}|}^2}
                     {{|g^{V}_{LL}|}^2 + {|g^{V}_{LR}|^2}} \right) 
   { = \frac{3}{4} \;\mbox{(SM)}} \\
  { \eta} &\propto& \Re(g^{V}_{LL} g^{S*}_{RR} + \cdots ) 
  {   = 0 \;\mbox{(SM)}}. 
\end{eqnarray}

The spin polarization-dependent Michel parameters are:
\begin{eqnarray}
  { \xi} &\simeq&  -\left( \frac{ {|g^{V}_{LL}| - 3|g^{V}_{LR}|}^2}
                     {{|g^{V}_{LL}|}^2 + {|g^{V}_{LR}|^2}} \right)
  {           =   -1 \;\mbox{(SM)}} \\
  { \delta} &\simeq& \frac{3}{4}\left( \frac{ {|g^{V}_{LL}|}^2}
                     {{|g^{V}_{LL}|}^2 + 3{|g^{V}_{LR}|^2}} \right)
  {           = \frac{3}{4} \;\mbox{(SM).}}
\end{eqnarray}

\subsection{Michel Parameter measurements}

There are updated results from 
LEP~\cite{ref:MichelALEPH,ref:MichelL3,ref:MichelOPAL,ref:MichelDELPHI}
and CLEO~\cite{ref:MichelCLEO}.
The world averages, compiled for TAU'98~\cite{ref:MichelStahl},
assuming lepton universality, are:
\begin{eqnarray}
\rho_\tau           &=& 0.750\pm 0.011 \q (SM = 3/4) \\
\eta_\tau           &=& 0.048\pm 0.035 \q (SM = 0)   \\
\xi_\tau            &=& 0.988\pm 0.029 \q (SM = 1)   \\
\xi_\tau\delta_\tau &=& 0.735\pm 0.020 \q (SM = 3/4).
\end{eqnarray}


The tau Michel parameter measurements are now
{\it precision} physics, although they are still
far from the precision obtained with muons~\cite{ref:PDG98}.
They are {\it consistent} with being entirely
$V-A$ in structure (left-handed vector couplings).
They strongly limit the probability of right-handed 
$\tau$ couplings to the weak charged current $P^\tau_R$;
for example, CLEO~\cite{ref:MichelMoritz}
sets the the limit $P^\tau_R < 0.044$ at 90\% CL.
However, for left-handed $\tau$ couplings,
it is currently not possible to distinguish 
between scalar, vector, and tensor contributions.
Independent information 
(\eg, from the cross section $\sigma(\nu_\tau e^-\to \tau^- \nu_e$))
is needed to distinguish between the possible
left-handed $\tau$ couplings.

The limit on right-handed couplings
can be interpreted in terms of limits on 
right-handed $W_R$ bosons.
In Left- Right symmetric models~\cite{ref:beg},
two charged boson mass eigenstates $M_1$, $M_2$ mix
to give the ``light'' $W_L$ of the SM, and a heavy $W_R$. 
The parameters in these models are
$\alpha = M(W_1)/M(W_2)$, and the mixing angle
$\zeta = $ mixing angle; both are zero in the SM.
The heavy right-handed $W^\pm_R$
will contribute to the decay of the tau,
interfering with the left-handed $W^-$,
and producing deviations from the Standard Model values
for the Michel parameters $\rho$ and $\xi$.
CLEO~\cite{ref:MichelMoritz} obtains limits on
$\alpha$ and $\zeta$ in these models,
shown in Fig.~\ref{fig:WLR}.
For mixing angle $\zeta = 0$, they obtain
$M_R > 304\mbox{ GeV/c}^2$ at 90 \% CL, and 
for free mixing angle $\zeta$, they obtain
$M_2 > 260\mbox{ GeV/c}^2$ at 90 \% CL.

\FIGURE[!htb]{
\epsfig{figure=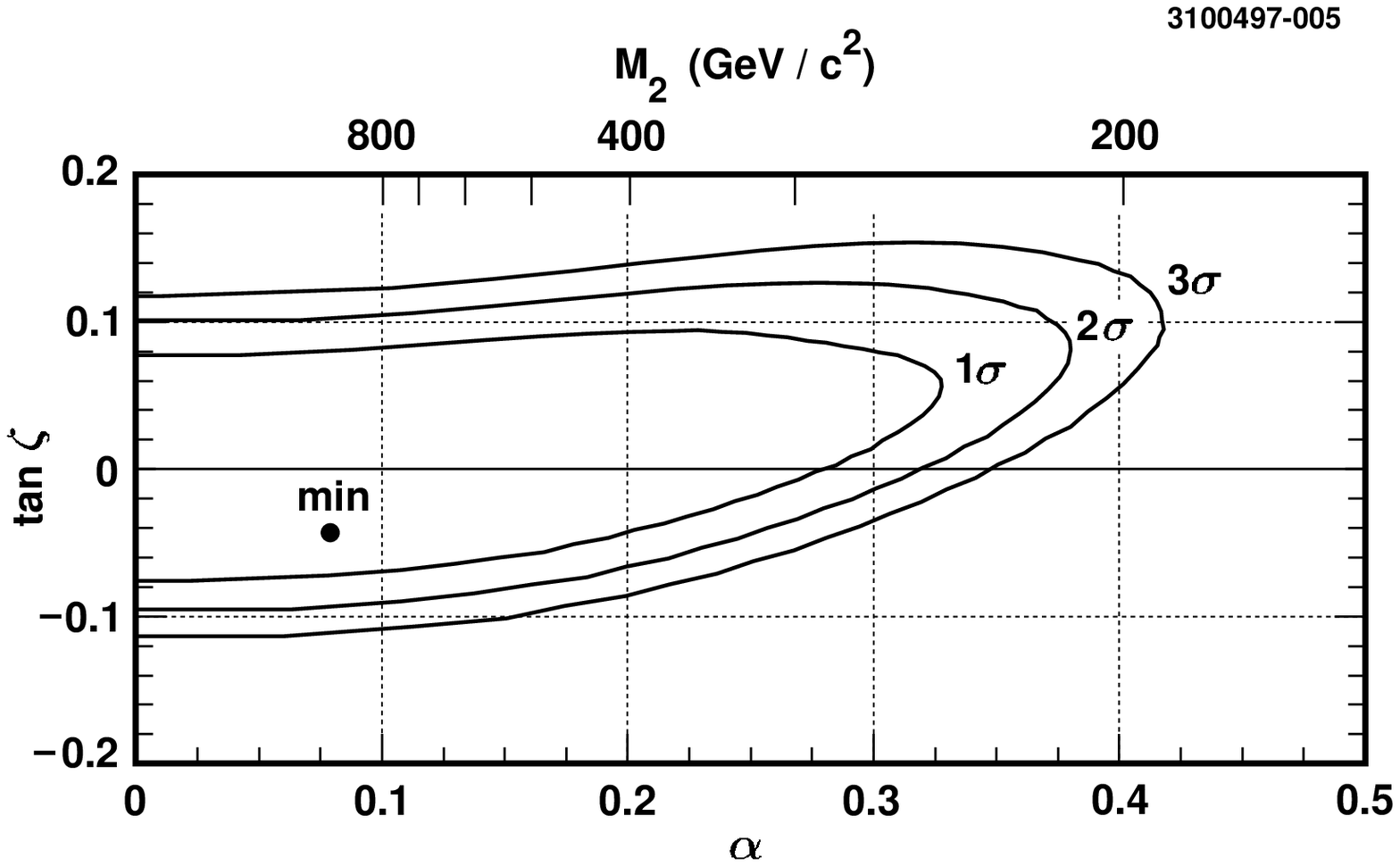,width=6.0cm}
\caption{Limits on the mass ratio $\alpha$ and the mixing angle
      $\zeta$ of a left-right symmetric model
      {\protect\cite{ref:MichelMoritz}}.}
\label{fig:WLR}
}

The consistency of the Michel parameters with SM predictions
permits a limit to be set on the mass of a 
(scalar) charged Higgs boson, in the context
of the MSSM~\cite{ref:Stahl}. For the $\eta$ parameter,
the dependency is:
\begin{equation}
\eta \approx - \;\;m_{\tau} m_{\mu} tan^{2}\beta/ (2 m^{2}_{H}),
\end{equation}
with similar formulas for ${ \xi}$ and ${ \xi\delta}$.
Using the world average measurements of 
$\eta$,  $\xi$, and $\xi\delta$ combined, one obtains:
\begin{equation}
    m_{H^\pm} > 2.5 \tan\beta\mbox{ GeV/c}^2 \q\hbox{at 90\% CL}.
\end{equation}
This is competitive with other direct and indirect
search limits (shown in Fig.~\ref{fig:chiggs})
for  two-Higgs-doublet mixing angles $\tan\beta \gsim 30$.

\FIGURE[!htb]{
\epsfig{figure=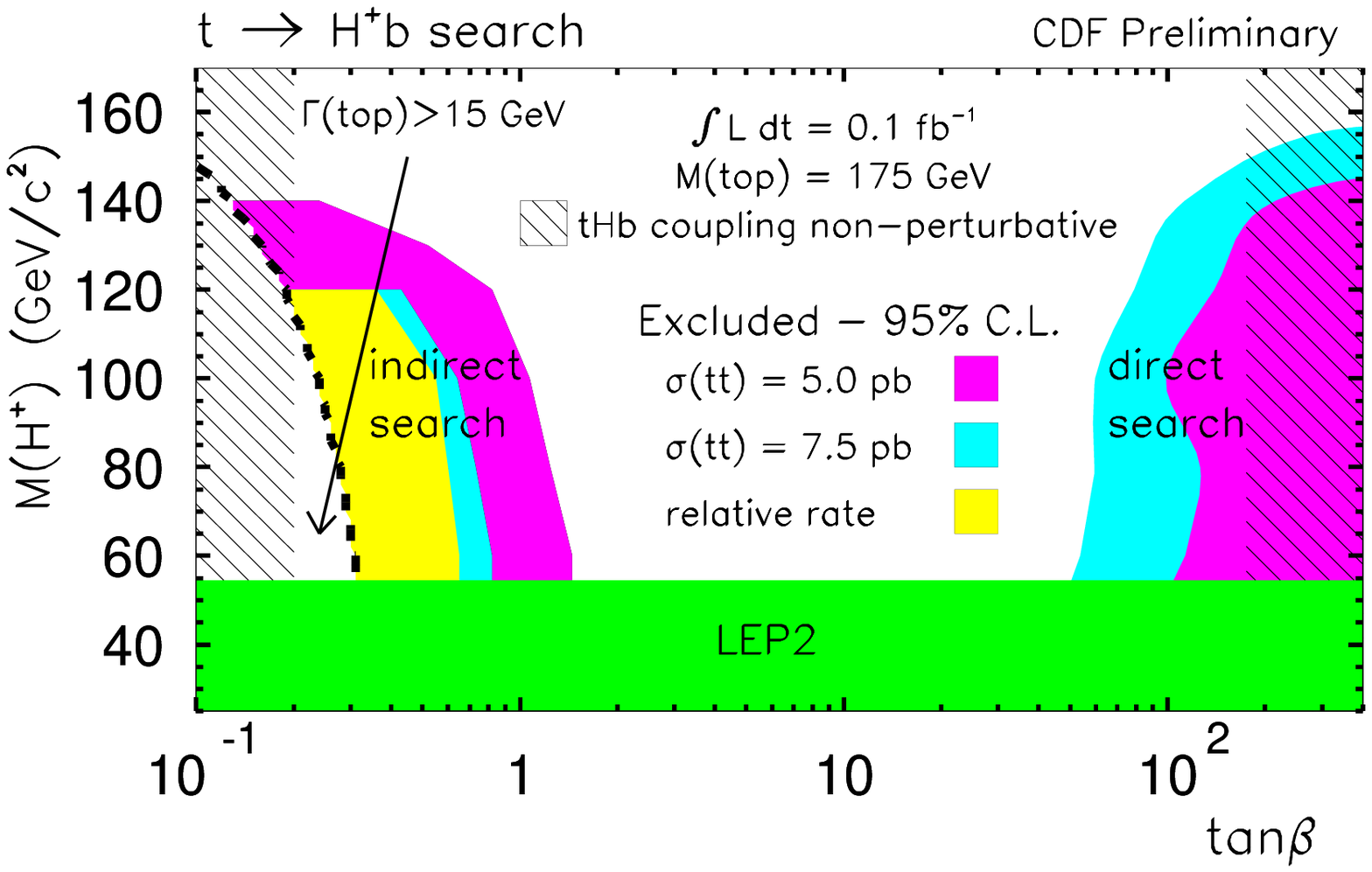,width=6.0cm}
\caption{Direct and indirect
search limits for charged Higgs mass versus $\tan\beta$.}
\label{fig:chiggs}
}

DELPHI has taken the analysis one step further,
by probing {\it derivative} terms in he interaction
Lagrangian (beyond the Michel ansatz).
DELPHI measures~\cite{ref:MichelDELPHI} the 
anomalous tensor coupling $\kappa$
by analyzing the tau leptonic decays 
with the usual Michel parameters fixed to their
SM values. They measure
\begin{equation}
\kappa = -0.029\pm0.036\pm0.018,
\end{equation}
in agreement with the SM expectation of $\kappa = 0$.

Once again, many extensions to the SM predict deviations
of these parameters from their SM values.
It is thus worth improving the precision of these measurements,
to push the limits on contributions from
charged Higgs, right-handed $W$'s, and other
anomalous couplings.

\section{$W\to \tau\nu$}

The strength of the weak charged current coupling 
to the $\tau$ can also be measured in $\tau$ production
from real $W$ decays.

\subsection{$W\to \tau\nu$ at LEP II}

All four LEP II experiments use the reaction
$e^+e^-\to W^+W^-$  to measure the ratio of rates
$(W\to \tau\nu)$ : $(W\to \mu\nu)$ : $(W\to e\nu)$.
The results are summarized~\cite{ref:LEPIItaunu}
in Fig.~\ref{fig:LEPIItaunu}.
There is excellent consistency between experiments,
final state leptons, and SM predictions.
Charged current universality is confirmed to 4.0\%\ 
via these measurements.

\FIGURE[!htb]{
\epsfig{figure=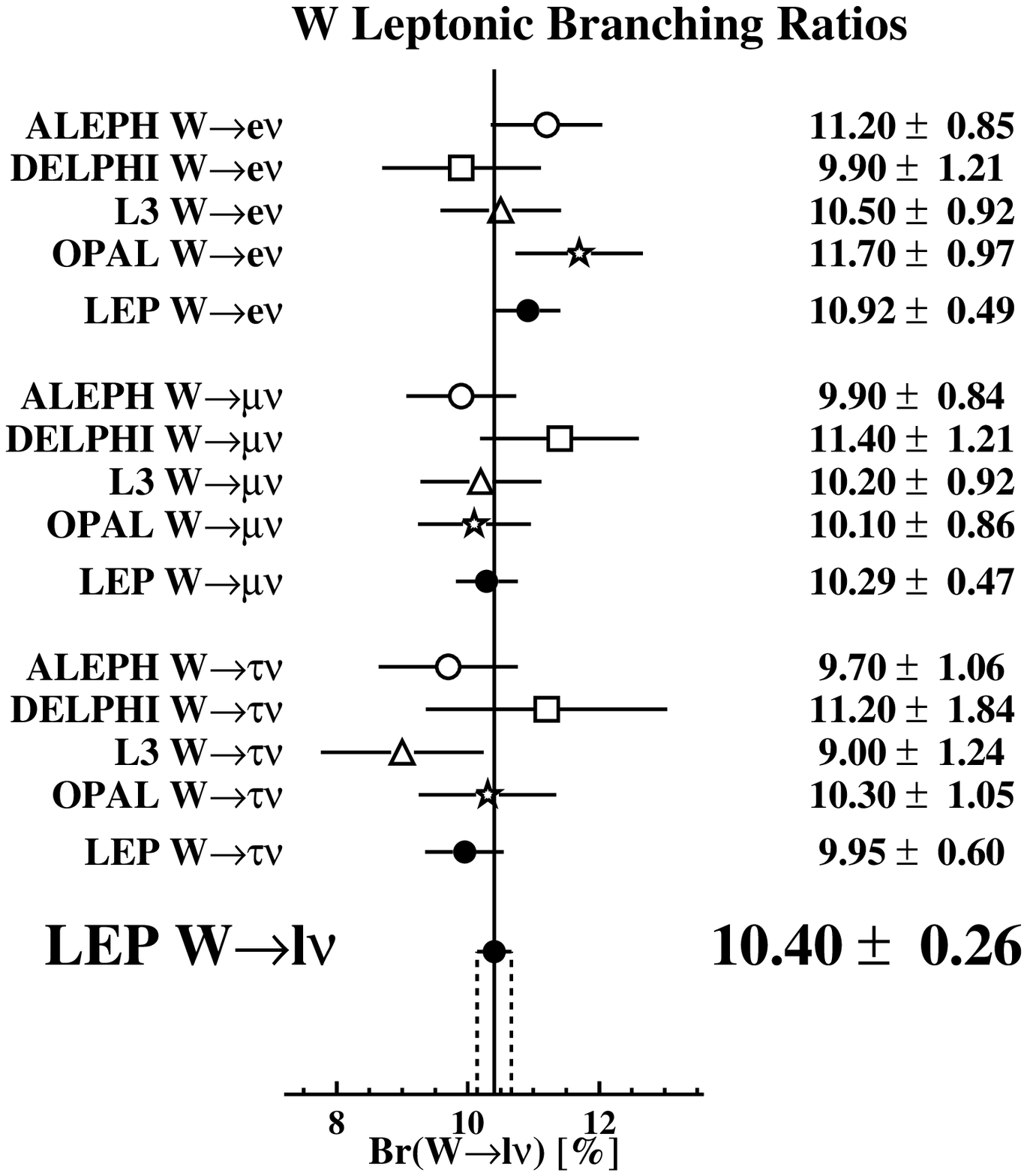,width=6.0cm}
\caption{
Branching fractions $\BR(W\to \ell\nu_\ell)$
measured by the four LEP II experiments
{\protect\cite{ref:LEPIItaunu}}.}
\label{fig:LEPIItaunu}
}

\subsection{$W\to \tau\nu$ from $p\bar{p}$}

Real $W$ bosons are also produced at $p\bar{p}$
colliders, via the reaction
$p\bar{p} \to W^\pm X$, $W\to \ell\nu$.
There are measurements of the coupling ratio
$g_\tau/g_e$ from UA1, UA2, CDF, and D0.
The world average~\cite{ref:pbarp}
is $g_\tau/g_e = 1.003\pm 0.025$,
confirming charged current universality at the 2.5\% level.
As seen in Section~\ref{s-univ} above,
charged current universality 
is tested in tau leptonic decays to 0.25\%.


\section{$Z^0$ Couplings}
\label{s-Z0coup}

The weak neutral current couplings
of the tau are directly measured in
tau pair production via $e^+e^-\to Z^0\to \tau^+\tau^-$.
All four LEP experiments, and SLD,
measure a large number of relevant observables.

\subsection{$R_\tau$ and $A_{FB}$}

The LEP experiments measure the 
ratio 
$R_\tau = \Gamma(Z\to\hbox{hadrons})/\Gamma(Z\to\tau\tau)$,
and the forward-backward asymmetry
$A_{FB}(Z^0\to\tau^+\tau^-)$.
The LEP averages for these quantities~\cite{ref:LEP_Rl}
and for the analogous quantities for the light leptons,
are shown in Fig.~\ref{fig:LEP_Rl}.
All three lepton species have values consistent
with each other and with the SM prediction
(assuming universality of the weak neutral current). 
In particular, the equality
$R_e = R_\mu = R_\tau$ is tested
to a precision of 0.3\%.

\FIGURE[!htb]{
\epsfig{figure=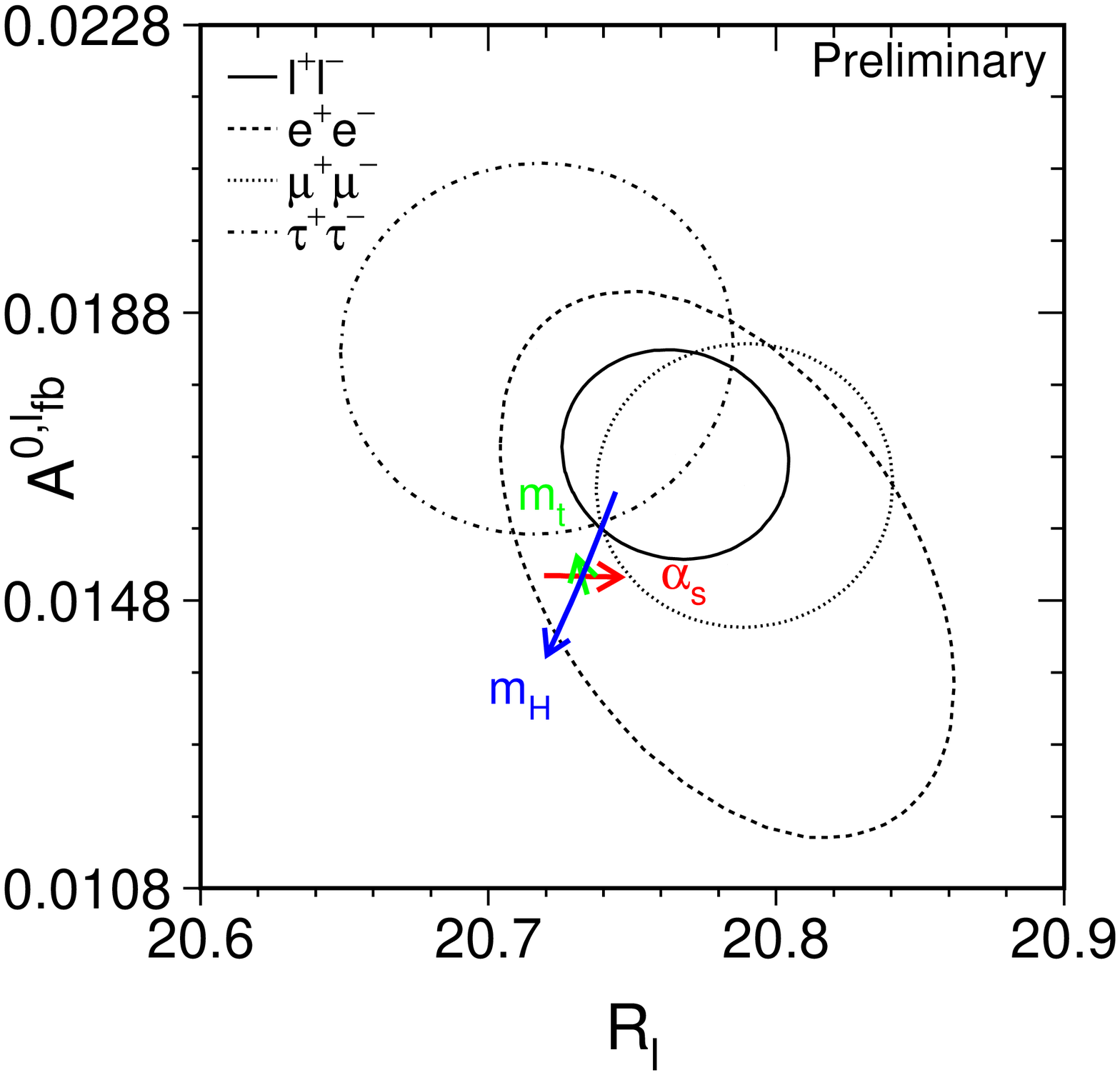,width=6.0cm}
\caption{Measurements of $R_\tau$ and $A_{FB}$
for $Z^0\to\ell^+\ell^-$, from LEP
{\protect\cite{ref:LEPgvga}}.}
\label{fig:LEP_Rl}
}


\subsection{$\tau$ polarization at $Z^0$}

All four LEP experiments measure the tau polarization
$P_\tau(\cos\theta)$
as a function of the $\tau$ production angle $\theta$,
using the decay modes $e\nu\nu$, $\mu\nu\nu$, 
$\pi\nu$, $\rho\nu$, and $3\pi\nu$.
From the measured $P_\tau(\cos\theta)$ distributions
(see example in Fig.~\ref{fig:Ptaucosth}),
they extract the asymmetry parameters
\begin{equation}
\calA_\ell \equiv (2 g_v^\ell g_a^\ell) /((g_v^\ell)^2 + (g_a^\ell)^2)
\end{equation}
for $\ell = e$ and $\tau$.
A summary of the results from LEP~\cite{ref:Ptausum}
is also shown in Fig.~\ref{fig:Ptaucosth}.
The world average results are
\begin{equation}
\calA_\tau = (14.31\pm 0.45)\%; \quad  \calA_e = (14.79\pm 0.51)\%.
\end{equation}

\FIGURE[!htb]{
            \epsfig{figure=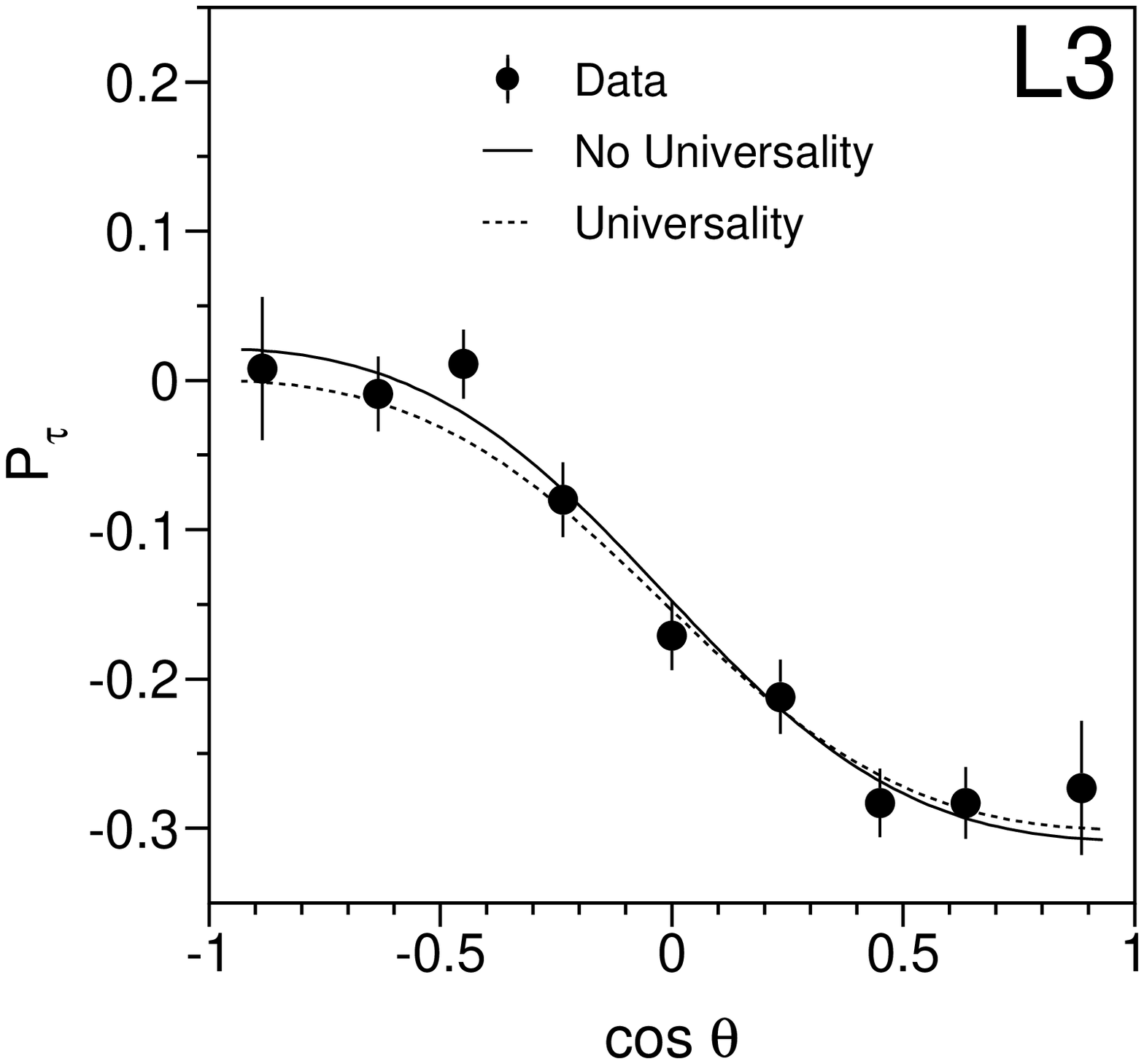,width=3.25cm}
            \hspace*{0.2cm}
            \epsfig{figure=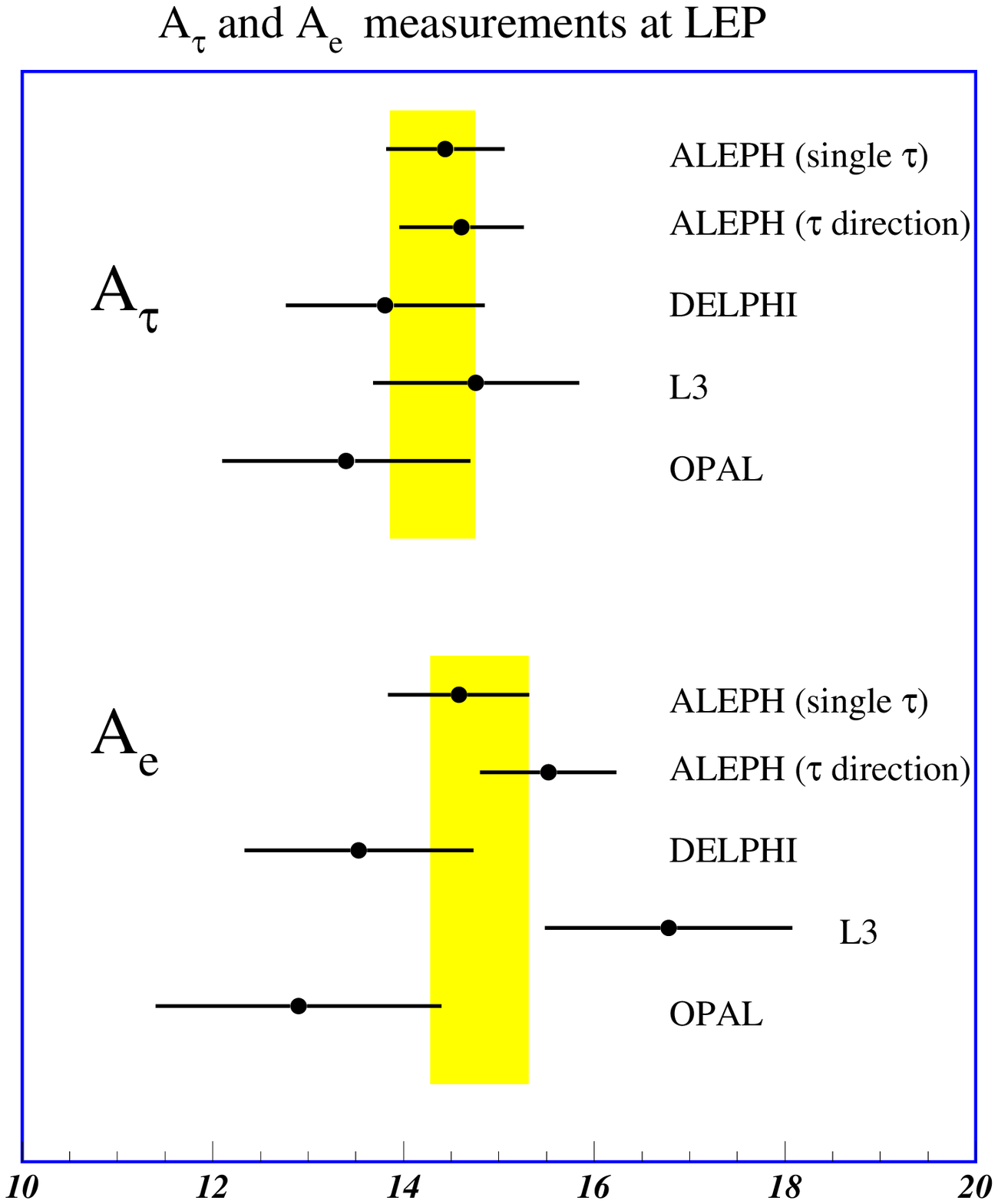,width=3.0cm}
\caption{Left: Measurement of $P_\tau(\cos\theta)$ from L3
{\protect\cite{ref:PtauL3}};
Right: summary of the results from LEP on $\calA_\tau$
and $\calA_e$ from tau polarization
{\protect\cite{ref:Ptausum}}.}
\label{fig:Ptaucosth}
}
            

\subsection{$A_{LR}^\tau(\cos\theta)$ from SLD}

SLD measures the tau polarization
as a function of the $\tau$ production angle $\theta$,
separately for left- and right-handed beam electron polarizations
at the SLC~\cite{ref:SLDALR}.
These are shown in Fig.~\ref{fig:SLDALR}.
From these measurements, they form the asymmetry 
$A_{LR}^\tau(\cos\theta)$.
This allows them to extract values for
$\calA_e$ and $\calA_\tau$ of relatively high precision,
despite low statistics.
They obtain~\cite{ref:SLDALR}
\begin{equation}
\calA_\tau = (14.2\pm 1.9)\%; \quad  \calA_e = (15.0\pm 0.7)\%.
\end{equation}

\FIGURE[!htb]{
\epsfig{figure=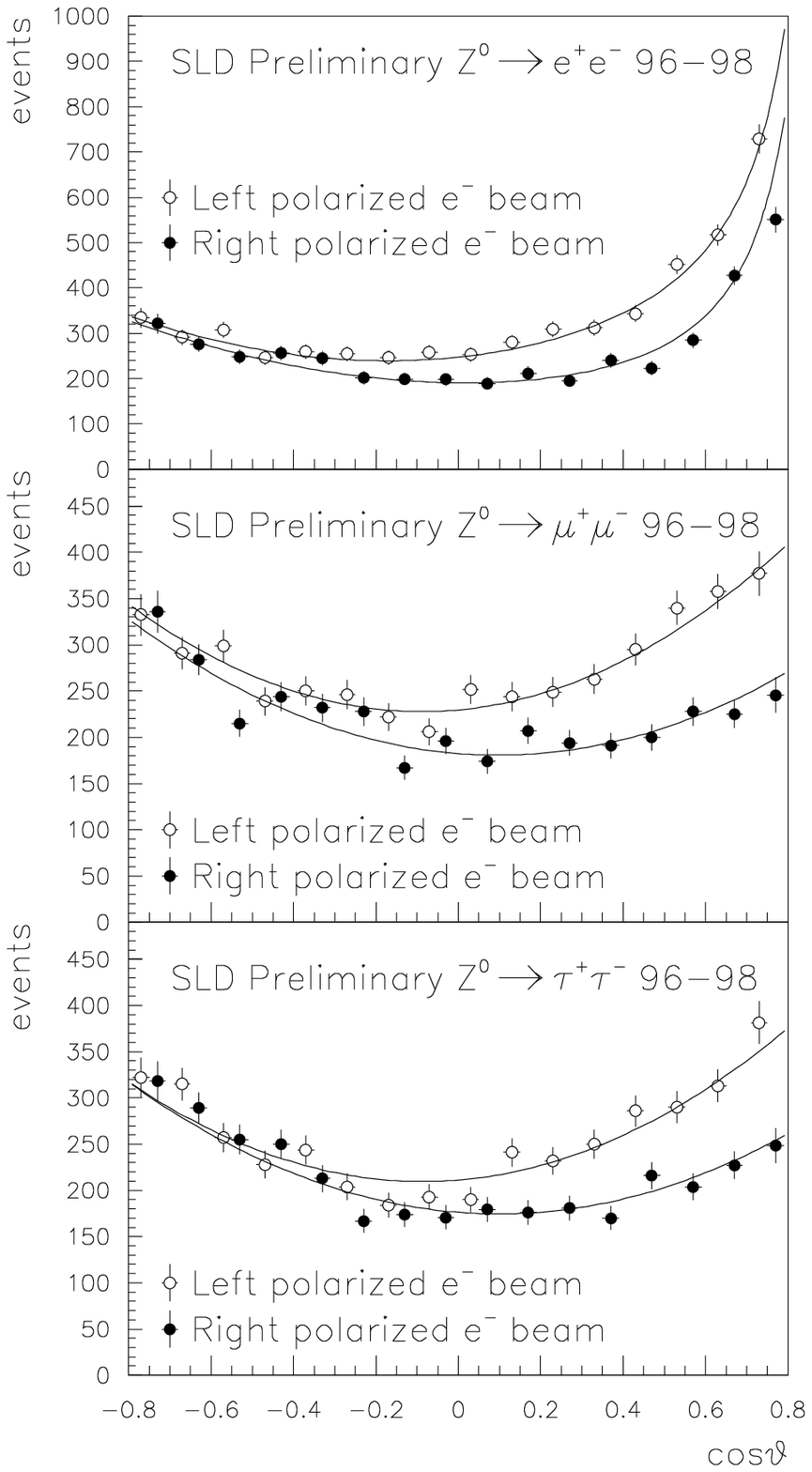,width=4.1cm}
\caption{The $\tau$ production angle distribution
 for left- and right-handed beam electron polarizations
 from SLD {\protect\cite{ref:SLDALR}}.}
\label{fig:SLDALR}
}

\subsection{NC Lepton Universality}

The measurements discussed in this section
can be combined to extract the vector and axialvector
weak neutral current coupling constants $g_v$ and $g_a$ 
for the tau (and the other leptons).
The results from LEP and SLD
for all three charged leptons is summarized~\cite{ref:LEPgvga}
in Fig.~\ref{fig:LEPgvga}.

There is fine agreement between experiments,
and all the leptonic couplings are consistent
with each other and with the SM prediction.
The latter depends on the SM Higgs mass, 
and it can be seen that a low-mass Higgs is favored.
Non-SM contributions, as measured
by the model-independent S and T parameters~\cite{ref:SandT}
are strongly constrained~\cite{ref:LEPgvga}.

\FIGURE[!htb]{
\epsfig{figure=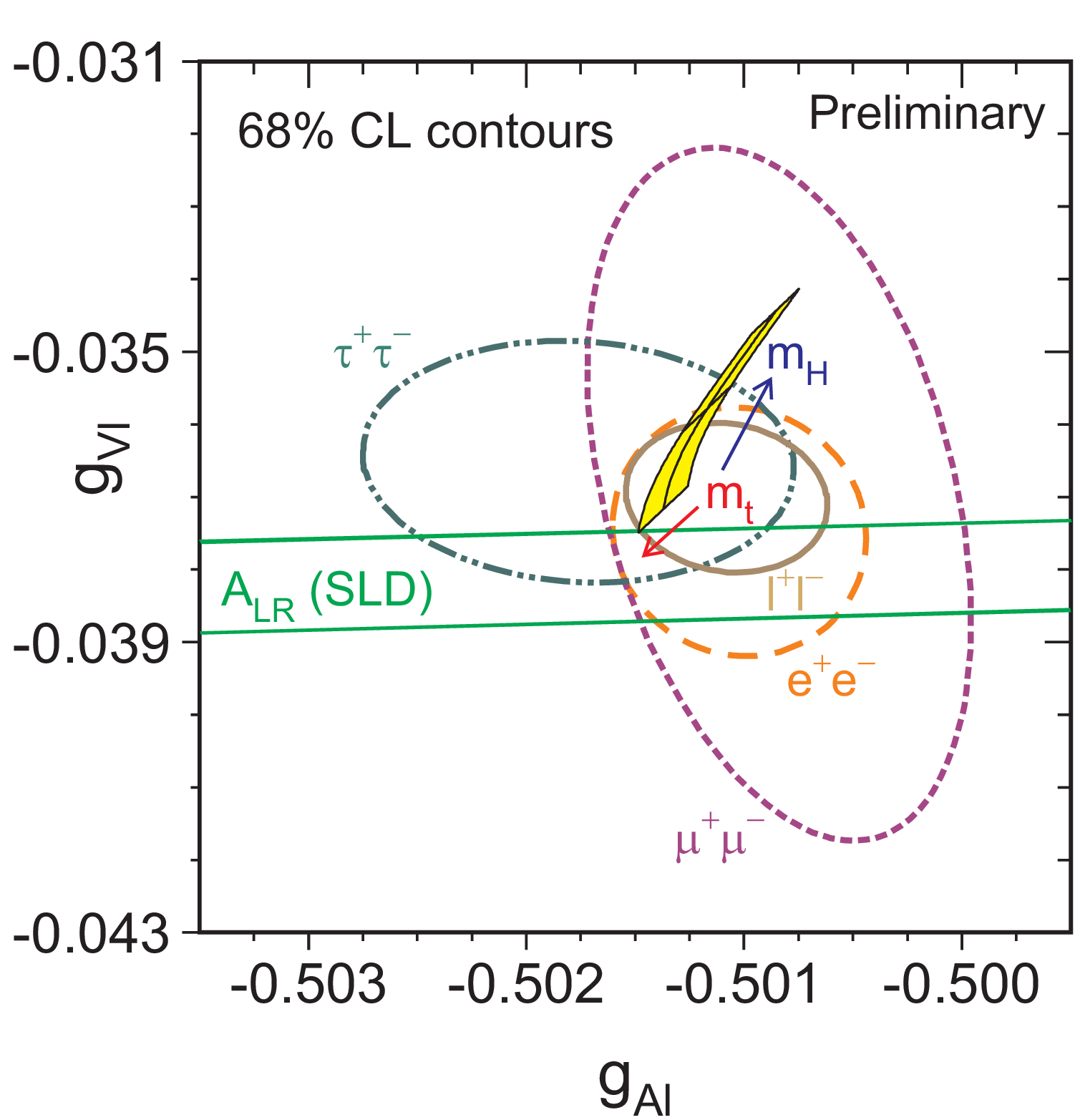,width=6.0cm}
\caption{Extracted values for the weak neutral current couplings
$g_v$ and $g_a$ for the leptons, from LEP and SLD
{\protect\cite{ref:LEPgvga}},
compared with SM predictions.}
\label{fig:LEPgvga}
}


Since all measurements are consistent with the SM predictions,
they can be used to extract the value of the
SM parameter $\sin^2\theta_{eff}$.
This can then be compared with the value,
and errors, for this parameter
obtained from studies of $Z^0\to q\bar{q}$.
This comparison~\cite{ref:LEPgvga}
is shown in Fig.~\ref{fig:sinW}.
We see that the measurement of $\cal_\tau$
provides one of the most precise methods
for obtaining $\sin^2\theta_{eff}$.
The LEP and SLD results are
completely consistent for the lepton measurements;
the LEP values for
$R_b$, $A_{FB}^b$, $A_{FB}^c$ pull the LEP average away from SLD,
but not very significantly so.

\FIGURE[!htb]{
\epsfig{figure=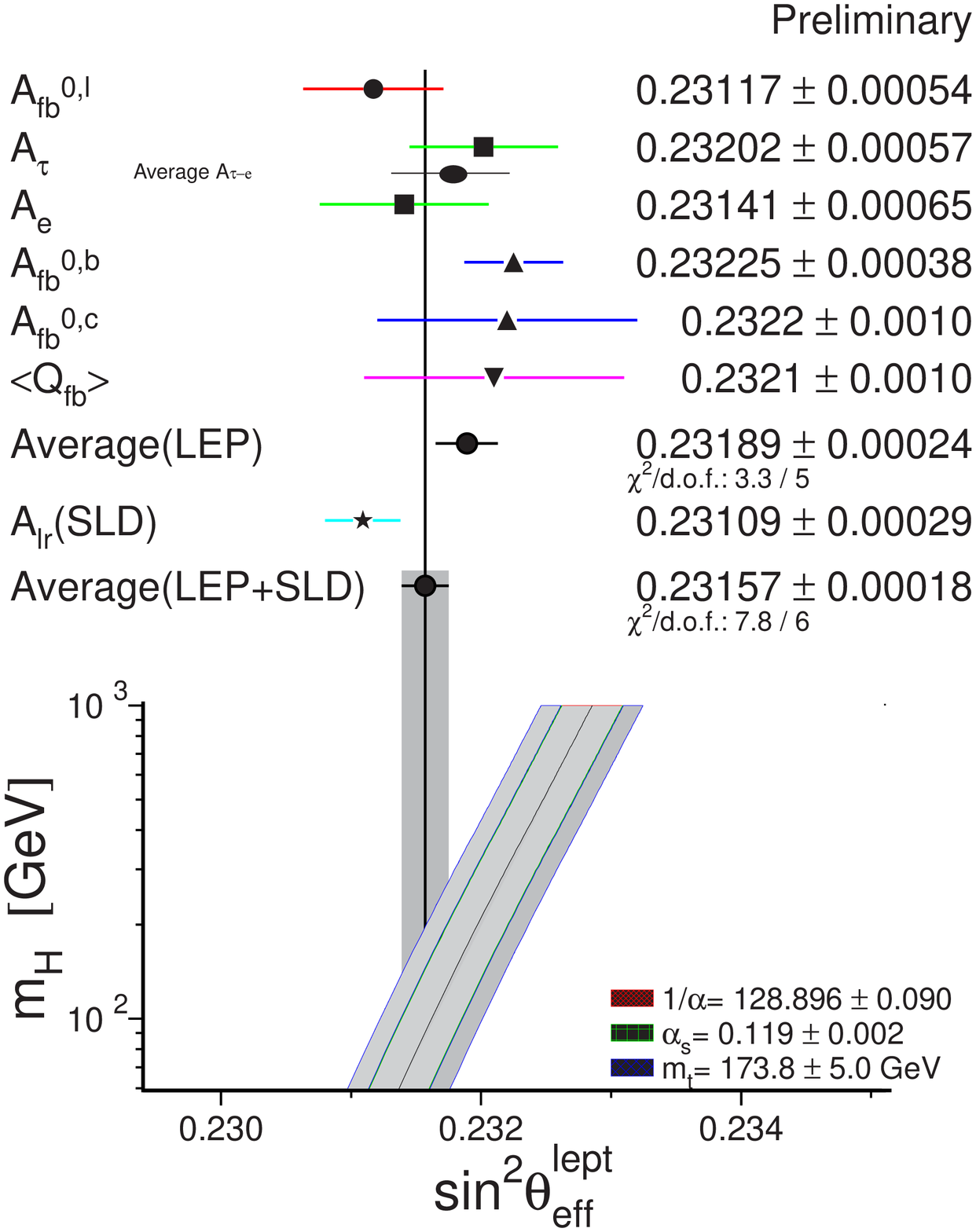,width=6.0cm}
\caption{Extracted values for $\sin^2\theta_{eff}$
from measurements at the $Z^0$ 
{\protect\cite{ref:LEPgvga}}.}
\label{fig:sinW}
}

\section{Dipole Moments}
\label{s-dipole}

The pure vector nature of the electromagnetic
couplings is modified due to radiative
corrections, which induce magnetic
dipole tensor couplings.
If the lepton or quark is composite, and if CP is violated,
electric dipole couplings also appear.

Analogous couplings also appear for the weak interactions.
In addition to the SM Lagrangian for $Z^0\tau\tau$,
which includes vector and axialvector couplings,
it is natural to consider extensions that
add tensor couplings, corresponding to 
weak electric and 
weak magnetic dipole moment couplings~\cite{ref:Bern}.
\begin{eqnarray}
{\cal L} &=& {\cal L}_{SM}
            + \frac{1}{2}\cdot\frac{eF_2^w(q^2)}{2m_\tau}
              \bar{\psi} \sigma^{\mu\nu}        \psi Z_{\mu\nu} \nonumber \\
         & &   - \frac{i}{2}\cdot\frac{eF_3^w(q^2)}{2m_\tau}
              \bar{\psi} \sigma^{\mu\nu}\gamma_5\psi Z_{\mu\nu}.
\end{eqnarray}
Here, $\psi$ is the quantum field of the tau,
$Z_{\mu\nu}$ is the $Z^0$ field strength tensor,
and $F_2^w(q^2)$ and $F_3^w(q^2)$ are the weak magnetic
and weak electric form factors, respectively.
The anomalous weak magnetic moment,
and the CP-violating weak electric dipole moment,
of the tau are:
\begin{equation}
a_\tau^w \equiv F_2^w(m_Z^2), \q\q
d_\tau^w \equiv \frac{e F_3^w(m_Z^2)}{2m_\tau}
\end{equation}

Predictions for the values of these
weak dipole moments, in the SM and beyond, 
are~\cite{ref:PredDipole,ref:Zalite}:
\begin{eqnarray}
  a_\tau^W &=& -(2.1 + 0.6i)\times 10^{-6} \hbox{(SM)} \\
           && \to 10^{-5} \q\hbox{(MSSM)} \\
           && \to 10^{-3} \q\hbox{(composite)} \\
  d_\tau^W &=& 3\times 10^{-37} \hbox{e$\cdot$cm} \q\hbox{(SM-CKM)} \\
           && \to \hbox{few}\q \times 10^{-20} \q\hbox{(MSSM, LQ)}.
\end{eqnarray}

Weak magnetic dipole couplings 
produce parity-odd azimuthal asymmetries~\cite{ref:azasym}.
For example, in $\tau^+\to\pi^+\nubar_\tau$,
the expectation value for 
$<\hat{p}_{\tau^+} \times \hat{p}_{e^+} \cdot \hat{p}_{\pi^+}>$
is proportional to $a_\tau^W$.
If it is anomalously large, it would be measurable at LEP.
L3 has measured this azimuthal asymmetry using
$\tau\to \pi\nu$ and $\rho\nu$.
Seeing no significant asymmetry, it sets the limits~\cite{ref:PredDipole}
\begin{eqnarray}
  Re(a_\tau^W) &=& (0.0\pm 1.6 \pm 2.3)\times 10^{-3} \\
  Im(a_\tau^W) &=& (-1.0\pm 3.6 \pm 4.3)\times 10^{-3} 
\end{eqnarray}

\subsection{CP violating Weak-Electric Dipole Moment}

A non-zero weak electric dipole moment (weak EDM) of the tau
would be evidence for both substructure and CP violation
in the lepton sector.
It would induce modifications to the spin structure
in $e^+e^-\to Z^0\to \tau^+\tau^-$~\cite{ref:Bern}.
The subsequent tau decays can be used to analyze
the spins of both taus in an event,
and seach for CP-odd spin polarizations and correlations.
These also take the form of triple product observables
which are CP-odd. 

A set of optimized CP-violating
observables have deen defined~\cite{ref:Bern},
and have been measured by the LEP experiments~\cite{ref:CPLEP},
using most tau decays ($\ell,\pi,\rho,a_1$)
as spin analyzers.
Simulated spectra, illustrating the effect
for non-zero weak EDM, are shown in Fig.~\ref{fig:WEDMspec}.
The measurements of Re($d_\tau^W$), Im($d_\tau^W$)
from LEP are shown in Fig.~\ref{fig:WEDMmeas},
and the limits from the combined data are~\cite{ref:Zalite}:
\begin{eqnarray}
|Re(d_\tau^W)| &<& 3.0\times 10^{-18} e\cdot cm \\
|Im(d_\tau^W)| &<& 9.2\times 10^{-18} e\cdot cm \\
|d_\tau^W|     &<& 9.4\times 10^{-18} e\cdot cm
\end{eqnarray}

\FIGURE[!htb]{
\epsfig{figure=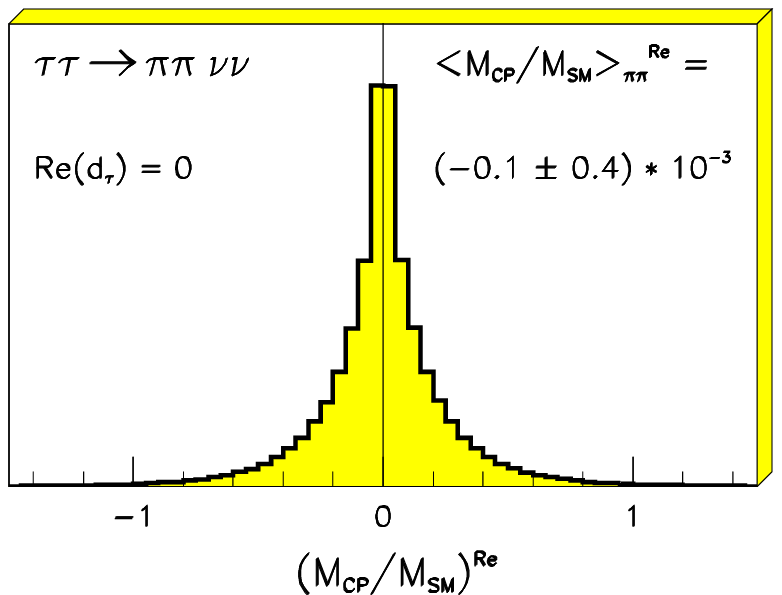,width=3.0cm}
            \hspace*{0.25cm}
            \epsfig{figure=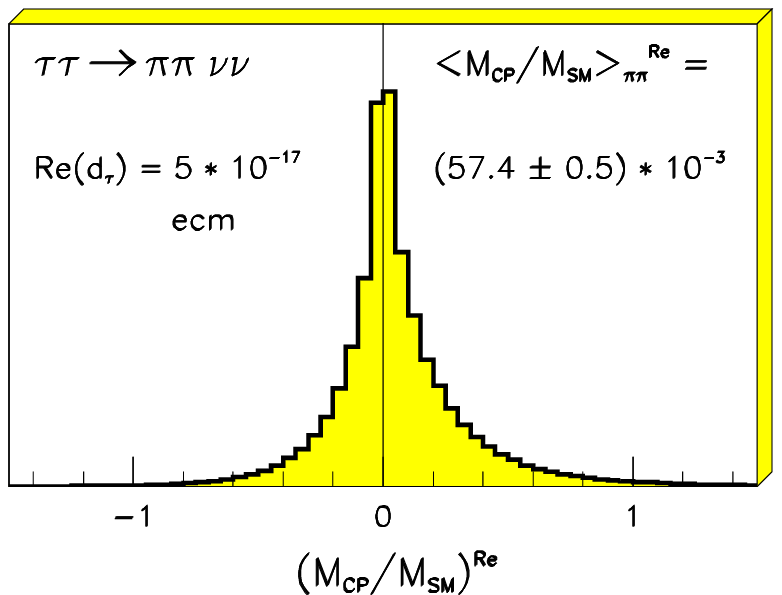,width=3.0cm}
\caption{Simulation of the distribtion of
optimal CP-violating observables in $Z^0\to \tau^+\tau^-$,
for the case of no WEDM (left), and for non-zero
WEDM (right) {\protect\cite{ref:Zalite}}.}
\label{fig:WEDMspec}
}

\FIGURE[!htb]{\epsfig{figure=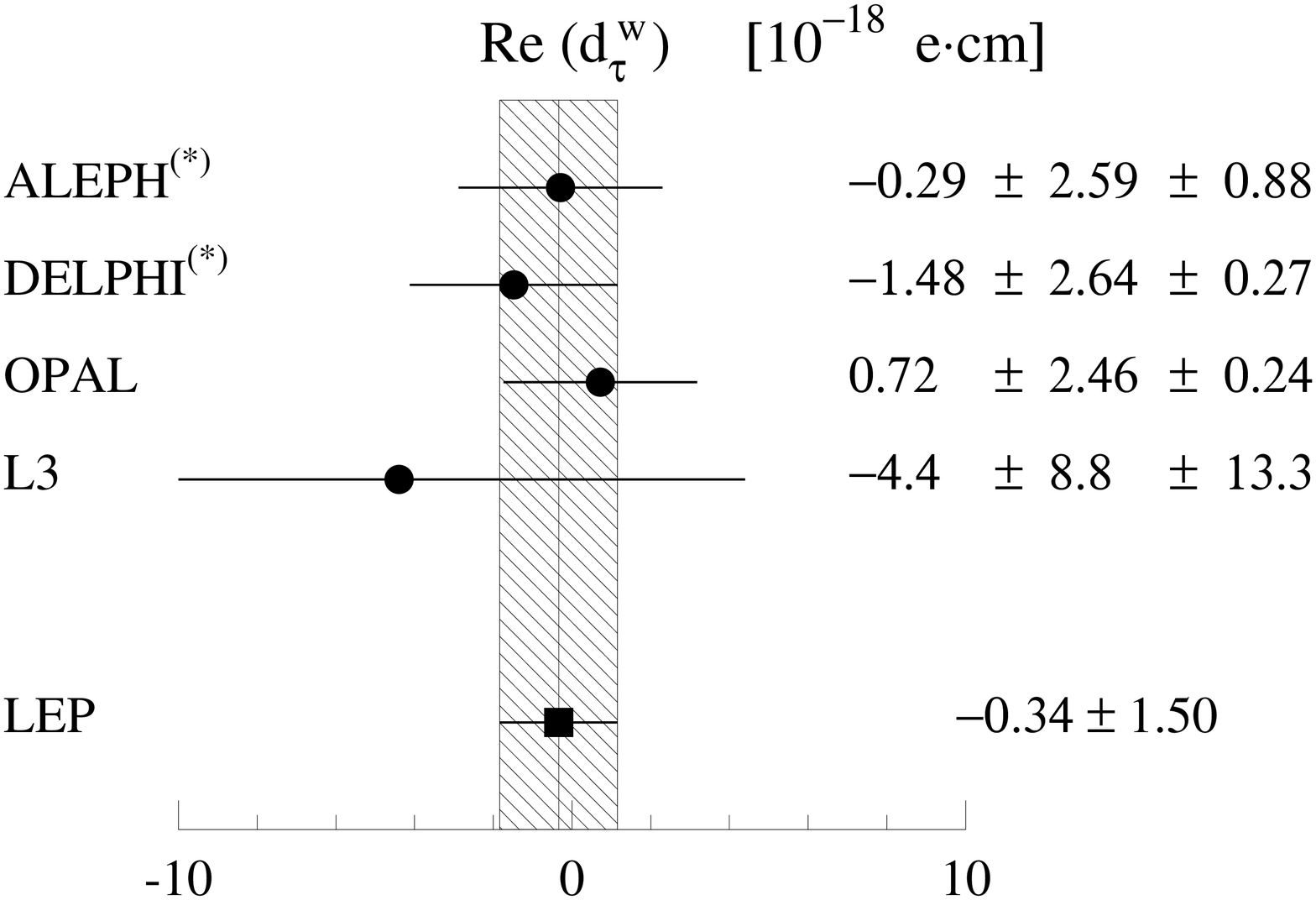,width=3.0cm}
            \hspace*{0.25cm}
            \epsfig{figure=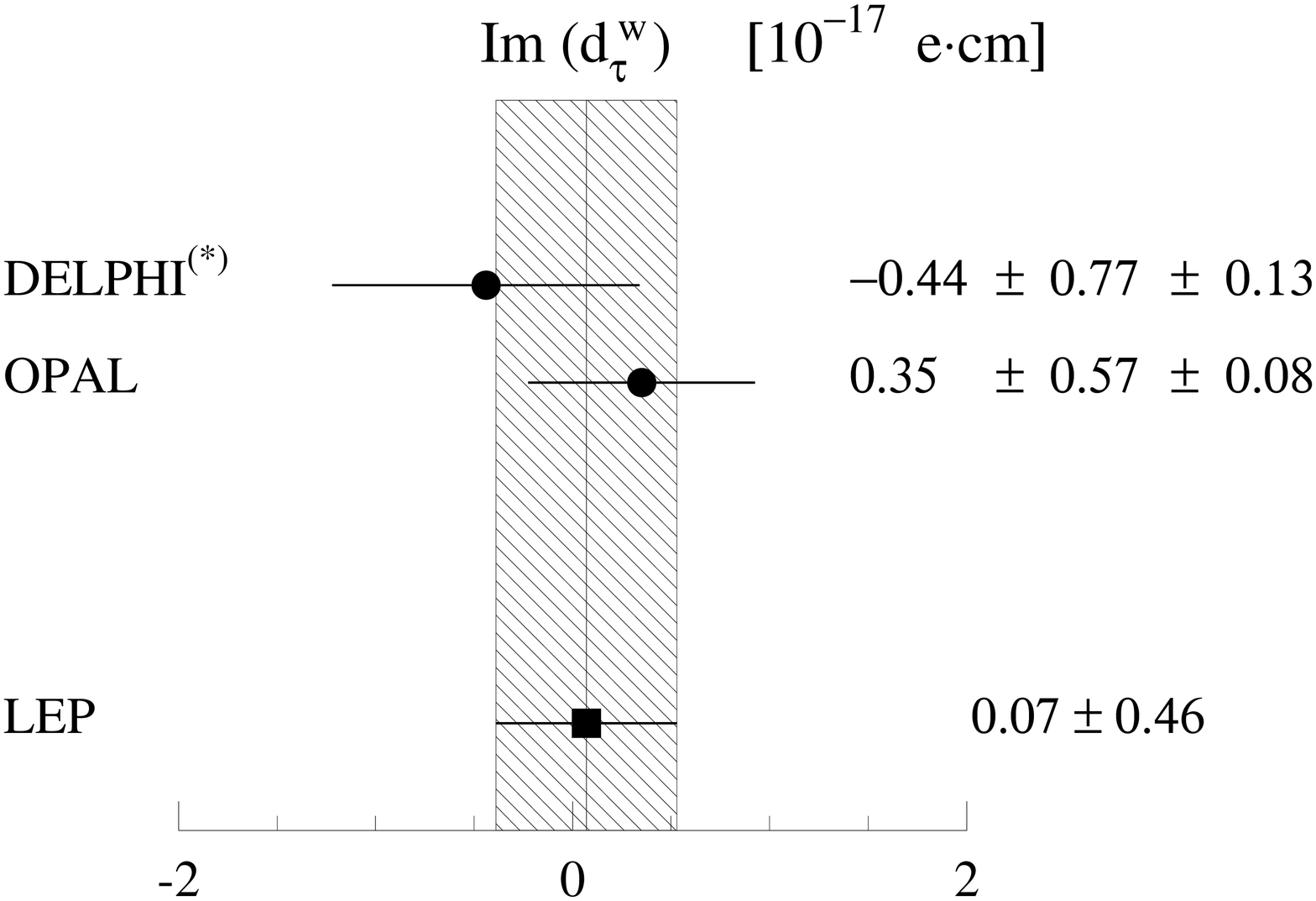,width=3.0cm}
\caption{Summary of recent measurements of 
the weak electric dipole moment of the tau
{\protect\cite{ref:Zalite}}.}
\label{fig:WEDMmeas}
}


\subsection{Weak Dipole Moments: SLD}

The electron beam longitudinal polarization
available at the SLC collider enhances the 
ability of the SLD detector to measure
the weak dipole moments, especially Im($d_\tau^W$). 
They do an unbinned likelihood 
fit to the full event kinematics,
using tau pairs which decay to $(\ell, \pi, \rho)$.
This allows them to measure the real and imaginary parts
of both weak dipole moments, and set the limits~\cite{ref:WDMSLD}:
\begin{eqnarray}
  Re(a_\tau^W) &<& 2.47 \times 10^{-3} \\
  Im(a_\tau^W) &<& 1.25 \times 10^{-3} \\
  Re(d_\tau^W) &<& 1.35 \times 10^{-17} e\cdot cm \\
  Im(d_\tau^W) &<& 0.87 \times 10^{-17} e\cdot cm
\end{eqnarray}
which are quite competitive with the LEP averages,
despite much smaller statistics.


\subsection{EM dipole moments}

Despite the dominance of the $Z^0$ over the virtual photon
at LEP I, the electromagnetic dipole moments
can be measured using radiative events,
$e^+e^- \to Z^0 \to \tau^+\tau^- \gamma$.
Anomalously large electromagnetic dipole moments
will produce an excess of events with a high energy photon,
away from both the beam $e^+$ and $\tau^+$ momentum axes~\cite{ref:EDMrad}.

L3~\cite{ref:EDML3}
and OPAL~\cite{ref:EDMOPAL}
compare the observed spectra in
$E_\gamma$ vs $\cos\theta_{\gamma}$ for radiated photons
to predictions from the SM with the addition of
anomalously large electromagnetic dipole moments, and
set limits on $a_\tau^\gamma$ and $d_\tau^\gamma$.
The L3 spectra are shown in Fig.~\ref{fig:EDML3}.
The resulting limits are~\cite{ref:EDMTaylor}:
\begin{eqnarray}
  |a_\tau^\gamma| &<& 0.06 \quad (SM: \frac{\alpha}{2\pi} = 0.011) \\
  |d_\tau^\gamma| &<& 3.1\times 10^{-16} \quad (SM: 0 (\CPv)).
\end{eqnarray}
For comparison, the limit on the electric dipole moment
of the electron is 
$|d_e^\gamma| < 5\times 10^{-25} e\cdot cm$~\cite{ref:PDG98}.

\FIGURE[!htb]{
\epsfig{figure=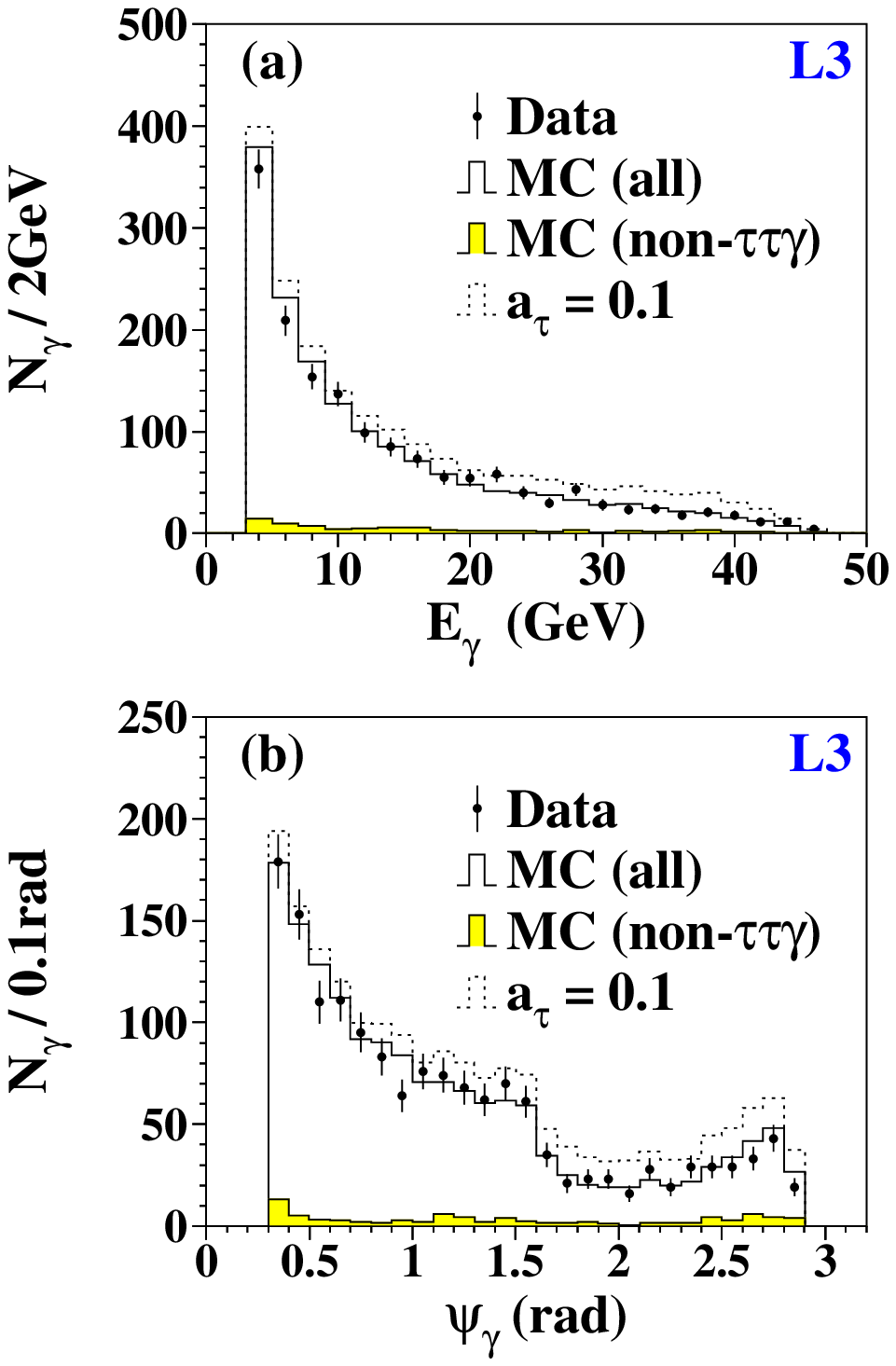,width=4.5cm}
\caption{Spectra of $E_\gamma$ and $\cos\theta_{\gamma}$
from L3 $\tau^+\tau^-\gamma$ events, 
compared with MC predictions with and without
an anomalously large magnetic dipole moment
{\protect\cite{ref:EDML3}}.}
\label{fig:EDML3}
}

\section{Other searches for new couplings}

Searches have also been made for other
non-SM currents such as the flavor-changing
neutral currents (FCNC) $\tau\lra e$ and
$\tau\lra\mu$ in neutrinoless tau decay,
and $\nu_\tau\lra\nu_e$ and $\nu_\tau\lra\nu_\mu$
in neutrino oscillation experiments.

\subsection{Neutrinoless tau decay}

All known tau decays proceed via the weak charged
current: $\tau^-\to\nu_\tau W^-$.
Flavor changing neutral current decays
such as $\tau^-\to e^- X^0$ and $\tau^-\to \mu^- X^0$,
where $X^0$ is some neutral current such as
the photon, $Z^0$, or some new current,
violate Lepton Flavor conservation.
LFV decays include:
        $\tau^- \to \ell^-\gamma$, $\ell^-\ell^+\ell^-$ 
        $Z^0 \to \tau^- e^+$, $\tau^-\mu^+$ 
        $\tau^- \to \ell^- M^0$, $\ell^- P_1^+ P_2^-$.
Here, $\ell$ is $e$ or $\mu$,
$M^0$ is a neutral meson, and $P^\pm$ is a
charged pseudoscalar meson.

Another class of decays violate Lepton Number conservation,
as well. LNV decays include:
        $\tau^- \ell^+ P_1^- P_2^-$, and
$\bar{p} X^0$. The latter conserves $B-L$.

SUSY, GUTS, Left-Right symmetric models, and
superstring models all predict 
LFV, LNV, and violations of the universality
of the dominant current couplings~\cite{ref:nuless}
The effects are small,
of the order of $10^{-6}$ or smaller,
and are only now within the reach of experiment~\cite{ref:stroy}.

The most sensitive search for neutrinoless decays
has been by the CLEO Collaboration,
which searches for $\tau^\pm\to\mu^\pm\gamma$
with 12.6 million produced tau pairs,
and sets the limit~\cite{ref:CLEOmugamma}:
$\BR(\tau^\pm\to\mu^\pm\gamma) < 1.1\times 10^{-6}$
at 90\%\ CL,
which is in the range of model parameters
for some supersymetric models~\cite{ref:nuless}.

CLEO also searched for 28 different 
neutrinoless decay modes, using
4.4 million produced tau pairs~\cite{ref:CLEOnuless}.
The limits on the branching fractions
are on the order of few $\times 10^{-6}$
or greater.

The present bounds are approaching or reaching levels
where some model parameter spaces can be excluded.
The models can also be pushed above the present limits;
so we are already beginning to exclude such efforts.
The current limits will be improved 
by the B Factory experiments, which
will push below the $10^{-7}$;
they will be rare $\tau$ decay experiments.


\subsection{Neutrino oscillations}

If one or more of the three neutrino flavor eigenstates
($\nu_e$, $\nu_\mu$, $\nu_\tau$) have mass
and can couple to the others,
they will mix and induce neutrino oscillations,
or (effective) flavor changing neutral currents.

Evidence for neutrino oscillations comes from
several different experiments~\cite{ref:nuosc}.
If the solar, atmospheric, {\it and} LSND 
observations are {\it all} correct,
it seems to require a $4^{th}$ (sterile? very massive?) 
neutrino~\cite{ref:nuosc}.

Only the atmospheric neutrino anomaly,
in which a deficit of muon neutrinos is observed
from cosmic ray showers,
is likely to involve the $\nu_\tau$.
The Super-Kamiokande experiment sees evidence
for $\nu_\mu\to\nu_X$ oscillations~\cite{ref:superK},
where $\nu_X$ may be a $\nu_\tau$
or a sterile $4^{th}$ generation neutrino $\nu_s$; however,
some evidence favors $\nu_\mu\to\nu_\tau$ over
$\nu_\mu \to \nu_{sterile}$~\cite{ref:sterile}.
The deficit is consistent with maximal
neutrino mixing ($\sin^2 2\theta \sim 1$),
and mass-squared difference 
$\Delta m^2_{\mu x} \sim 10^{-2} \hbox{eV}^2$.

This observation has spawned a host of mid- and long-baseline
accelerator experiments, in which a $\nu_\mu$ neutrino beam
from pion decay travels some distance, allowing it to
oscillate into a neutrino of different flavor,
which is then detected by a detector capable of distinguishing
$\nu_\mu \to \mu X$ from $\nu_\mu \to \nu_\tau \to \tau X$.
Two mid-baseline experiments at CERN, 
CHORUS~\cite{ref:CHORUS} and NOMAD~\cite{ref:NOMAD}, 
have completed their run.
Having failed to observe the latter reaction,
they exclude $\nu_\mu \to \nu_\tau$ for 
$\Delta m^2 \gsim 40$ eV$^2$, $\sin^2 2\theta \gsim 2\times 10^{-4}$,
as shown in Fig.~\ref{fig:CHORUS}.

\FIGURE[!htb]{
\epsfig{figure=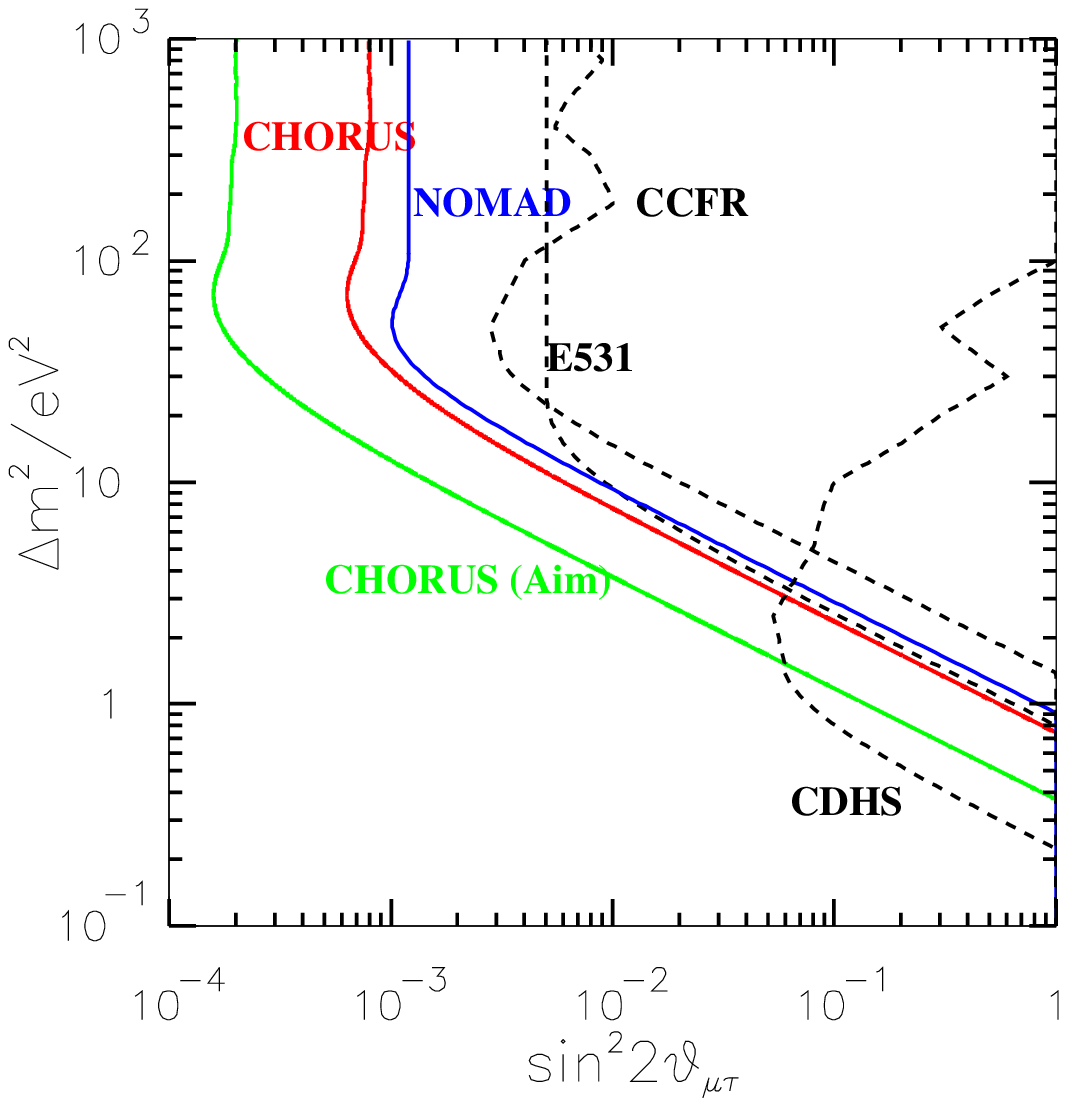,width=6.0cm}
\caption{Exclusion limits for 
$\nu_\mu \to \nu_\tau$ in the space of
$\Delta m_\nu^2$ versus $\sin^2 2\theta_{mix}$,
from recent mid-baseline experiments.
{\protect\cite{ref:CHORUS}}.}
\label{fig:CHORUS}
}

In order to reach the small $\Delta m^2$ suggested by
Super-K, a new generation of long-baseline experiments
are being prepared~\cite{ref:longbase},
including K2K, FNAL to MINOS, and CERN to Gran Sasso.
These experiments will probe the region down to
$\Delta m^2 \sim 10^{-3}$ ev$^2$,
$\sin^2 2\theta \gsim 10^{-1}$,
as illustrated in Fig.~\ref{fig:icarus}.
The understanding of these flavor changing neutrino couplings
is one of the major goals of particle physics in the next decade.

\FIGURE[!htb]{
\epsfig{figure=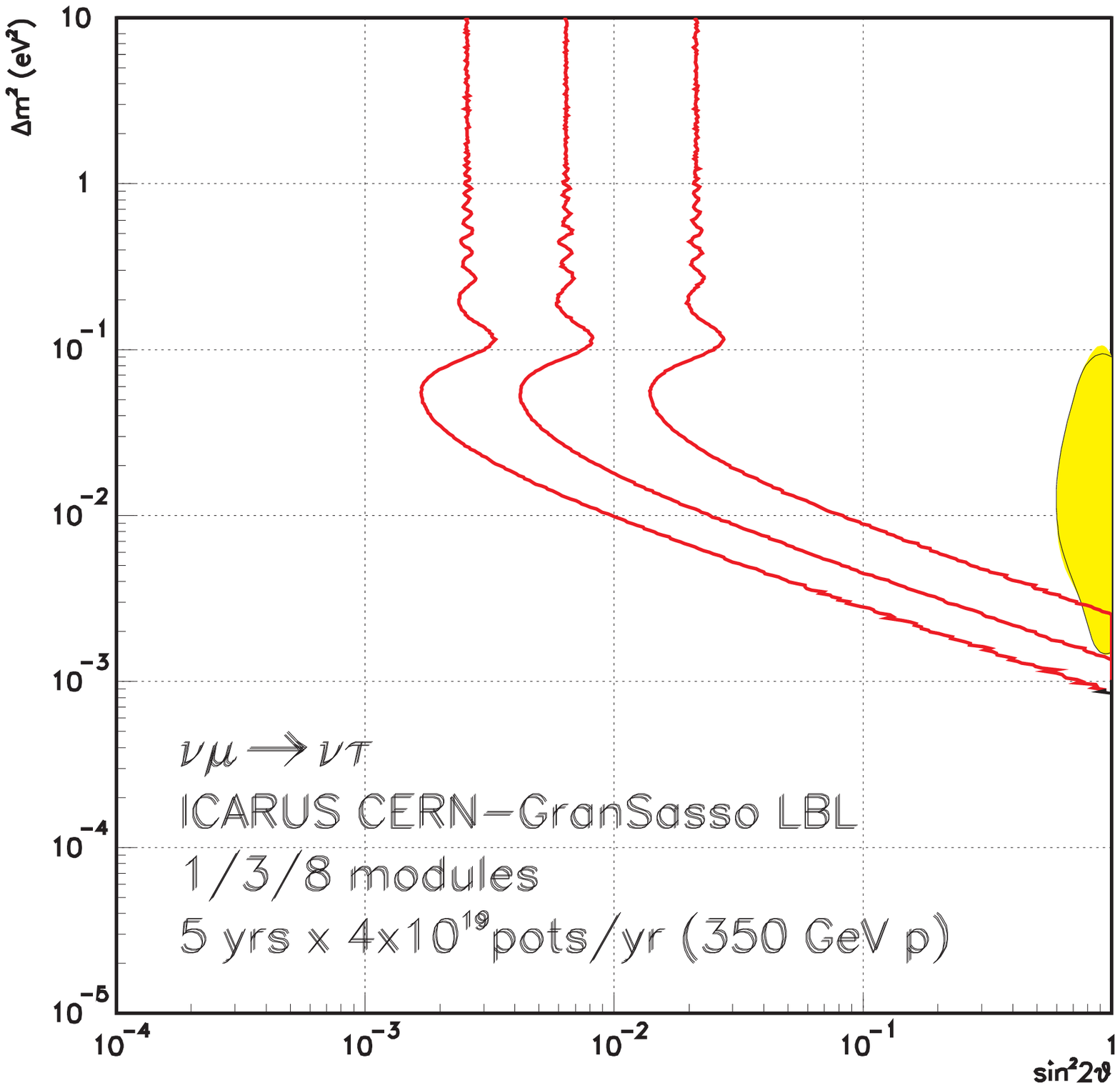,width=6.0cm}
\caption{Exclusion limits for 
$\nu_\mu \to \nu_\tau$ in the space of
$\Delta m_\nu^2$ versus $\sin^2 2\theta_{mix}$,
expected from the ICARUS long-baseline experiment.
The Kamiokande observation is shown in yellow.}
\label{fig:icarus}
}

\section{Limits on $m_{\nu_\tau}$}

Requiring the mass density of neutrinos to be
less than that required to over-close the universe
excludes stable neutrinos with masses larger
than 65 eV~\cite{ref:mcnulty}.
However, if neutrinos decay, they can evade that limit.
For lifetimes in the range of
$\sim$1 day $< \tau_{\nu_\tau} < \sim$ few years,
neutrino masses on the order of $\sim 5 < m_\nu < 20$ MeV
are allowed, where the upper bound comes from direct searches
in tau decays.

The technique that has yielded the best limits on
the $\nu_\tau$ mass in the tens-of-MeV region
is the study of the two-dimensional
mass and energy spectrum of the $n\pi$ final state
in $\tau\to (n\pi)^-\nu_\nu$, $n\ge 3$.
A deficit of events in the corner of $m_{n\pi}$, $E_{n\pi}$ space
indicates the recoil of a massive neutrino.
However, the spectrum is falling sharply there,
leading to very limited statistics;
and the spectral function governing that spectrum
is not precisely known.
The best limits obtained so far~\cite{ref:mcnulty}
are listed in Table~\ref{tab:mnutau}.

\TABLE[h]{
\label{tab:mnutau}
\caption{Summary of limits on $m_{\nu_\tau}$
         from $\tau\to (n\pi)^-\nu_\nu$, $n\ge 3$.}
  \begin{tabular}{llr} \hline
   ALEPH & $5\pi(\pi^0)$   & 22.3 MeV \\
   ALEPH & $3\pi$          & 30 MeV \\
   ALEPH & both            & 18.2 MeV \\
   OPAL  & $5\pi$          & 39.6 MeV \\
   DELPHI & $3\pi$         & 28 MeV \\
   OPAL   & $3\pi - vs - 3\pi$ & 35 MeV \\
   CLEO (98)  & $5\pi$, $3\pi2\pi^0$ & 30 MeV \\
   CLEO (98)  & $4\pi$               & 28 MeV \\ \hline
  \end{tabular}
}

It is interesting to note that the limits from CLEO~\cite{ref:CLEOmnu}
are not as tight as those from LEP, despite much larger
statistical samples. Indeed, there are 
many subtle issues involved in making these measurements,
regarding resolution, backgrounds, 
event migration, spectral functions,
and the fluctuations of low statistics.
The larger samples expected from the B Factory experiments
should help clarify the situation considerably,
and potentially improve the limits to the 10 MeV/c$^2$ range.

\section{Future prospects, and conclusions}

Experiments at LEP, SLD, and CLEO have produced
a wealth of rather precise measurements of the 
electroweak couplings, including limits
on a range of potential couplings beyond
the Standard Model ones.
But new physics may (hopefully) be just around the corner,
and higher precision in these these very fundamental
measurements may reveal it.
The fact that the tau is the heaviest known lepton,
free of uncertainties from non-perturbative physics,
makes it a particularly sensitive probe of new,
high mass scale physics.

The LEP $Z^0$ program is now over, but 
the B Factories now coming on line 
(CLEO III, BaBar, and Belle) will produce 
on the order of $10^7$ $\tau^+\tau^-$ per year.
This will permit a wealth of new measurements,
including:
 rare decays ($7\pi\nu$, $\eta \pi\pi\nu$, \etc);
 forbidden ($\nu$-less) decay (limits?);
 $m_{\nu_\tau}$ to $\lsim 10$ MeV;
 greater precision on universality tests;
 greater precision on  Michel Parameters,
 probing Higgs and $W_R$ couplings;
 weak and EM dipole moments, CP violation; and
 deeper studies of low-mass meson dynamics.
We may also see the observation of $\nu_\mu\lra\nu_\tau$ 
oscillations in the long-baseline experiments now 
in preparation.

We can expect continued progress in $\tau$ physics
in the coming years, and maybe (someday) some surprises!


\acknowledgments

The author acknowledges many fruitful conversations
with M.~Davier.
Thanks go also to the members of the LEP and SLD collaborations
who provided their latest results.
Finally, the author thanks the organizers of Heavy Flavors 8
for a very enjoyable, fruitful, and well-run conference.

\end{document}